\documentclass[manuscript]{acmart}
\usepackage{soul}  
\usepackage{color} 
\usepackage{xspace}
\usepackage{listings}

\newcommand{\eg}{{\it e.g.,\ }}

\newcommand{\ie}{{\it i.e.,\ }}

\definecolor{activegold}{RGB}{255,193,61}

\definecolor{lightorange}{RGB}{230, 170, 50}
\definecolor{lightgreen}{RGB}{121,210,121}
\definecolor{lightteal}{RGB}{121,199,210}
\definecolor{lightblue}{RGB}{100,212,239}
\definecolor{lightpurple}{RGB}{153,102,255}
\definecolor{lightred}{RGB}{245, 132, 120}
\definecolor{red}{RGB}{178,34,34}
\definecolor{gray}{RGB}{166,166,166}

\newcommand{\Toolname}[1]{\textit{CoNewsReader}}
\newcommand{\yuan}[1]{{\color{black} #1}}
\newcommand{\ldd}[1]{{\leavevmode\color{black} #1}}

\newcommand{\revise}[1]{{\color{black} #1}}

\newcommand{\ky}[1]{{\color{black} #1}}

\newcommand{\kangyu}[1]{{\color{black} #1}}

\newcommand{\cscwrr}[1]{{\color{black} #1}}
\newcommand{\cscwfr}[1]{{\color{black} #1}}
    
\AtBeginDocument{%
  \providecommand\BibTeX{{%
    \normalfont B\kern-0.5em{\scshape i\kern-0.25em b}\kern-0.8em\TeX}}}

\usepackage{multirow}
\usepackage{makecell}
\usepackage{placeins}
\usepackage{svg}

\setcopyright{acmlicensed}
\copyrightyear{2018}
\acmYear{2018}
\acmDOI{XXXXXXX.XXXXXXX}

\acmConference[Conference acronym 'XX]{Make sure to enter the correct
  conference title from your rights confirmation emai}{June 03--05,
  2018}{Woodstock, NY}
\acmISBN{978-1-4503-XXXX-X/18/06}

\begin{document}

\title{\Toolname{}: Supporting Comprehensive Understanding and Raising Critical Thoughts on Social Media News Through Comments}

\author{Kangyu Yuan}
\authornote{The majority of this work was completed while the author was studying at Sun Yat-sen University.}
\email{kyuanaf@connect.ust.hk}
\orcid{0009-0001-8460-9651}

\affiliation{%
  \institution{Sun Yat-sen University}
  \city{Zhuhai}
  \country{China}
}

\affiliation{%
  \institution{The Hong Kong University of Science and Technology}
  \city{Hong Kong}
  \country{China}
}

\author{Guanzheng Chen}
\email{22403048@zju.edu.cn}
\orcid{0000-0002-0152-9120}
\affiliation{%
  \institution{College of Education, Zhejiang University}
  \city{Hangzhou}
  \country{China}
}

\author{Sizhe Liang}
\email{liangszh5@mail2.sysu.edu.cn}
\orcid{0009-0001-0216-7395}
\affiliation{
 \institution{Sun Yat-sen University}
 \city{Zhuhai}
 \country{China}
}

\author{Hehai Lin}
\email{hlin709@connect.hkust-gz.edu.cn}
\orcid{0009-0009-2602-9059}
\affiliation{%
  \institution{The Hong Kong University of Science and Technology (Guangzhou)}
  \city{Guangzhou}
  \country{China}
}

\author{Qingyu Guo}
\email{qguoag@connect.ust.hk}
\orcid{0000-0002-8509-3933}
\affiliation{%
 \institution{The Hong Kong University of Science and Technology}
 \city{Hong Kong}
 \country{China}}

\author{Dingdong Liu}
\email{dliuak@connect.ust.hk}
\orcid{0000-0003-0985-0979}
\affiliation{%
  \institution{The Hong Kong University of Science and Technology}
  \city{Hong Kong}
  \country{China}
}

\author{Xiaojuan Ma}
\email{mxj@cse.ust.hk}
\orcid{0000-0002-9847-7784}
\affiliation{%
  \institution{The Hong Kong University of Science and Technology}
  \city{Hong Kong}
  \country{China}
}

\author{Zhenhui Peng}
\authornote{Corresponding author}
\email{pengzhh29@mail.sysu.edu.cn}
\orcid{0000-0002-5700-3136}
\affiliation{
 \institution{Sun Yat-sen University}
 \city{Zhuhai}
 \country{China}
 }

\renewcommand{\shortauthors}{}



\begin{teaserfigure}
\centering
\includegraphics[scale=0.27]{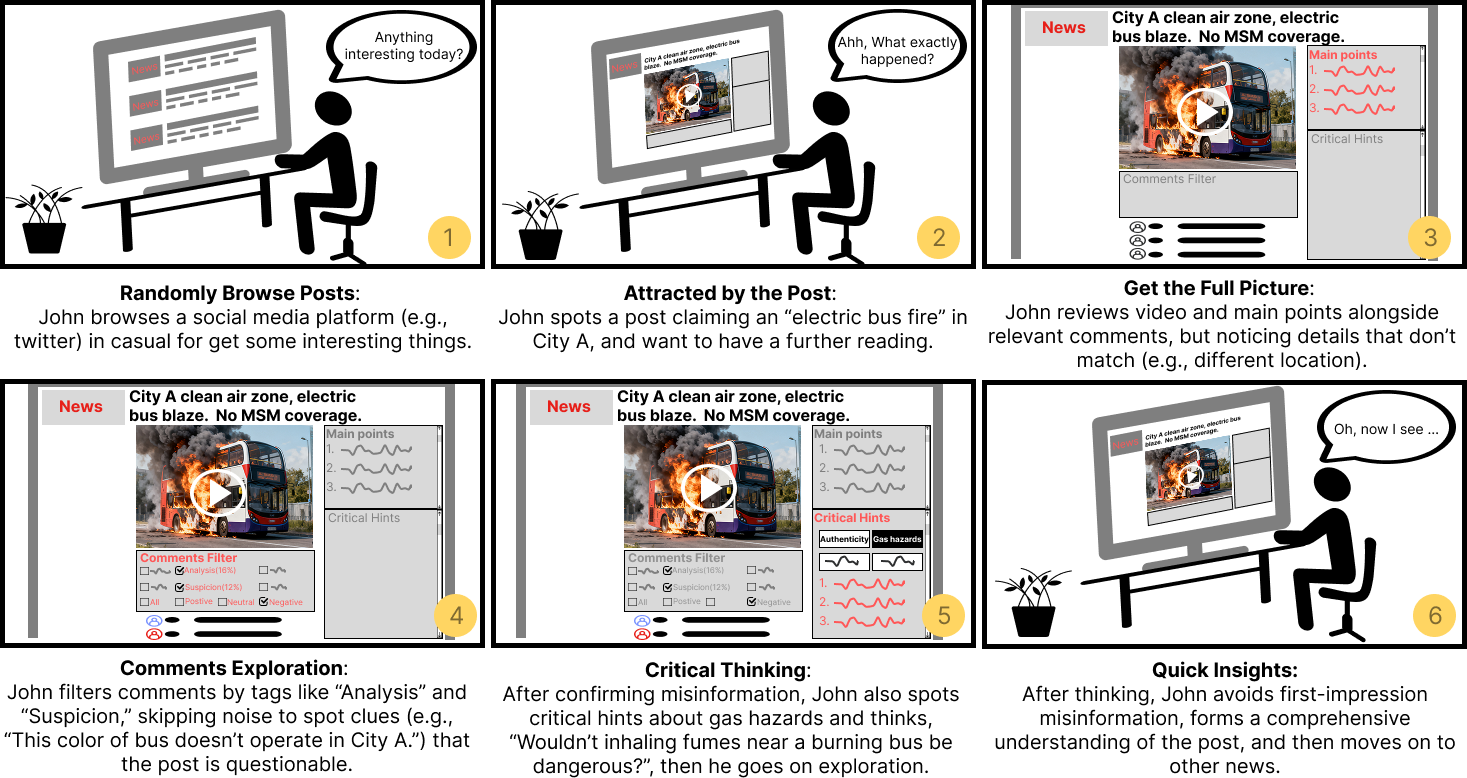}
\caption{General user scenario. A common social media news reading case reveals how readers use main points, critical-target comments filter, and critical thinking hints to uncover the misinformation and achieve a comprehensive understanding.}
\label{teaser}
\end{teaserfigure}

\begin{abstract} 
Critical news reading (CNR), which requires grasping the holistic ideas of and raising critical thoughts on the news, is beneficial yet challenging for \cscwfr{general} people who usually get information on \cscwfr{daily} social media. Comments under the news can aid CNR by providing complementary information and other readers' diverse and critical thoughts. However, it is under-investigated how to leverage these comments to support users in CNR. In this paper, we first derive user requirements for a comment-based CNR tool from literature and a formative study (N=12). Then, we develop \Toolname{}, a comment-based interactive CNR tool powered by a large language model. \Toolname{} supports users in grasping the news idea with complementary information from comments, filtering useful comments for CNR, and getting questions generated based on the comments to conduct critical thinking. Our within-subjects study \cscwfr{with 24 university students} indicates that compared to a baseline news reading interface in social media, participants with \Toolname{} have a more engaging CNR experience and perform better on comprehending the news and raising critical thoughts. We discuss design considerations for supporting reading tasks with user- and machine-generated content. 
\end{abstract}


\maketitle
\section{Introduction}

Reading ``we-media'' news (news and information collected, reported, analyzed, and disseminated by general citizens \cite{bowman2003we}) on social media has become a daily practice for many people \cite{SocialNetworkUsage}. 
Unlike traditional journalism which typically operates within established editorial frameworks, ``we-media'' content is created and shared by individuals with diverse motivations \cite{lacy2015defining}: from informing to persuading, advertising, or simply drawing public attention.
This citizen-led news ecosystem operates with less standardized editorial oversight or fact-checking processes, often resulting in content that presents partial perspectives of events and may contain strong personal opinions \cite{guo2013linking, jin2017news}.
These structural characteristics create significant challenges for news readers attempting to apply critical news reading (CNR), a demanding skill which requires readers to not only understand, analyze, and evaluate news content holistically but also approach it with appropriate skepticism \cite{kiili2022adolescents, nash2021constructing, sinatra2020evaluating}.
Effective CNR can help \cscwfr{general} readers distinguish plausible and credible information on social media, develop critical perspectives on news events, and apply these insights in their daily lives \cite{sinatra2020evaluating, nash2021constructing}.

The comment sections on social media platforms present a potential resource for addressing these challenges.
\cscwrr{We use comment as a general term to refer to user responses attached to a thread-starting post about a news item (\eg reply tweets), rather than only comments directly posted under the original news article on news websites, which functionally express opinions, stances, and analyses related to a news event.}
Through community participation, users share opinions, provide complementary information, and critique news content \cite{jones2019r, loo2019there}.
These diverse perspectives and additional context can aid readers in developing a more comprehensive understanding of news events and foster critical thought \cite{loo2019there, allagui2021ted}, which are two essential components of the CNR process.    
However, effectively utilizing these comments remains challenging.
Comment sections often contain overwhelming volumes of content with inconsistent quality \cite{diakopoulos2011towards}, while existing filtering mechanisms on platforms like Twitter and Weibo prioritize engagement metrics (\eg number of replies or likes) rather than supporting critical news reading \cite{jahanbakhsh2022putting}.
This gap between the potential value of comments for critical news reading and the limitations of current systems motivates our research toward proposing tools that help readers leverage comment sections for more effective critical news consumption.

In this paper, we propose \Toolname{} \footnote{\Toolname{} facilitates users to co-read news with others via their comments.}, an interactive tool that leverages comments to support critical news reading by facilitating information gathering, diverse perspectives evaluation, and critical thought development on social media.
\kangyu{Figure \ref{teaser} shows a common case of \cscwfr{general users} conduct social media news reading with \Toolname{}.}
Our approach builds upon existing tools for comment navigation and summarization \cite{faridani2010opinion, huh2022cocomix} and critical thinking support systems \cite{peng2022crebot, yuan2023critrainer}.
While Opinion Space \cite{faridani2010opinion} enables exploration of comments based on attitude and popularity, and CReBot \cite{peng2022crebot} provides in-situ critical thinking prompts for scientific reading, previous tools have not specifically addressed how comments can support the critical news reading process. 
Key questions remain about which types of comments are most valuable for CNR, how to model and integrate these comments into the reading process, and what the effectiveness and user experience of a comment-driven interactive CNR tool might be.

To address these questions, we develop a structured approach (3R workflow) to critical news reading based on established literature \cite{brown2015teaching, biringkanae2018use, dharma2013implementation}. 
Our 3R workflow guides readers through obtaining a comprehensive understanding of news content (\cscwfr{\textbf{R}ead content}), engaging with comments to understand diverse perspectives (\cscwfr{\textbf{R}ead comment}), and developing their own critical reflections (\textbf{R}eflect) (detailed in Section~\ref{sec:3r}).
To inform the design of our system, we conduct a formative study with \ky{12} university students to identify challenges in critical news reading and preferences for comment utilization. 
This leads to the development of a taxonomy of comments, including CNR-facilitative categories such as Information Enrichment, Personal Engagement, and Critical Reflection. 
We then implemented \Toolname{}, a tool that leverages this taxonomy to support each stage of the 3R workflow. 

We evaluate \Toolname{} through a within-subjects study with 24 participants who regularly engage with comments during news reading. 
Results demonstrate that participants using \Toolname{} develop a more comprehensive understanding of news content and generate more focused, diverse, logical, and solid critical reflections compared to a baseline Twitter-like interface. 
A complementary study with 8 participants who typically do not read comments \kangyu{and 2 participants in a daily social media news reading scenario} reveals that \Toolname{} increases their interest in comment exploration and supports critical reading processes.


In summary, the main contributions of this work are:
\begin{itemize}
    \item A taxonomy of useful comment types that support information gathering, perspective evaluation, and critical thought development in news reading contexts
    \ky{\item \Toolname{}, a novel system that helps users leverage comments to enhance critical news reading on social media}
    \item Empirical evidence demonstrating the effectiveness of comment-driven tools in improving news comprehension and critical thinking
    \ky{\item Design considerations for future news reading support tools addressing feature development, interaction design, and comment utilization}
\end{itemize}

\section{Related Work}

\subsection{Critical News Reading with Comments in Social Media}
Critical thinking is an essential skill for twenty-first-century education \cite{binkley2012defining, tiruneh2016systematic}, involving the ability to gather relevant information, assess source credibility, and raise critical questions \cite{lai2011critical, ennis1989critical, halpern1998teaching}. 
When applied to news consumption on social media, critical news reading (CNR) requires readers to analyze and evaluate information plausibility and credibility from diverse perspectives while forming well-supported personal opinions \cite{sinatra2020evaluating, kiili2022adolescents, nash2021constructing}.
The CNR process typically involves multiple stages. First, readers must comprehend the news content, understanding not only the event description but also grasping the publisher's intent and considering other readers' viewpoints \cite{bessu2018using, chaffee2014thinking, almoqbel2019understanding}.
Following comprehension, critical readers engage in deeper analysis, including distinguishing facts from opinions, examining hidden meanings, and ultimately forming well-reasoned conclusions \cite{chaffee2014thinking, wallace2003critical, kuta2008reading}. 
Common CNR practices include evaluating sources, seeking external information, examining headlines, and reading comment sections \cite{dykstra2019critical, mcclune2010critical, springer2015user}.

Comment sections can potentially serve as valuable resources for supporting CNR by providing diverse perspectives and additional context. 
Several studies have demonstrated how comments can enhance critical engagement with content.
For example, \citet{allagui2021ted} showed that engaging with TED comments enhanced students' critical thinking by exposing them to multiple perspectives.
While comments offer potential benefits for CNR, current studies and social media platforms present significant limitations in leveraging this potential. 
\textbf{First, a significant gap exist in the need for a taxonomy of comment types specifically organized according to their functional value in supporting different stages of the CNR process.}
Although researchers have extensively studied comment characteristics such as valence \cite{heinbach2018sleeper, hong2018will, waddell2020authentic, winter2015they}, civility \cite{anderson2014nasty, prochazka2018effects}, reasoning quality \cite{prochazka2018effects}, argument types \cite{winter2017user}, exemplification \cite{spence2017so}, and complexity \cite{comments_news_engagement_cscw2021}, these taxonomies were not specifically designed to support the CNR process on social media news platforms. 
Most existing comment categorizations focus on either content characteristics or discourse patterns rather than their functional value for critical reading.
Furthermore, studies like \citet{jones2019r} examined moderated forums with established community guidelines rather than general social media platforms, which contain more varied and potentially lower-quality comments.
\textbf{Second, there is still a lack of social media platforms that can organize comments based on their functional value for CNR and present them in ways that reduce cognitive load while highlighting diverse perspectives.}
A major challenge is that comment sections often contain overwhelming volumes of content \cite{lee2017news}.
Current platforms typically prioritize comments based on engagement metrics rather than their potential to support critical reading. 
These lead to a cognitive overload and making it difficult for readers to identify valuable perspectives amid the noise.
Research consistently shows that comments can substantially influence how readers perceive news content, hindering critical reading due to exemplification effects \cite{lee2017news}.
For instance, negatively-valenced comments, when appear in large volumes, have been shown to diminish the persuasive power of articles \cite{heinbach2018sleeper, winter2015they, winter2016s, winter2017user}, activate reader prejudices, and alter perceptions of reality \cite{lee2017user}.
They can also reduce perceived article credibility \cite{heinbach2018sleeper, waddell2020authentic}, issue importance, and relevant behavioral intentions \cite{spence2017so, ziegele2018socially}.
However, \citet{lee2017user} found that exposure to positive comments at the same time could help readers overcome this bias.
This effect reveals the importance to fully get access to diverse comments.

\ldd{
Our work addresses these gaps by: (1) developing a taxonomy of comment types specifically organized to support different stages of critical news reading; (2) designing and implementing a comment-driven system that integrates these comment types into a structured reading workflow; and (3) evaluating how such a system affects users' ability to comprehend news content critically and develop well-reasoned reflections.}

\subsection{News Reading Support Tools}
HCI researchers have explored a series of tools to support news reading activities in digital devices\cite{chen2023marvista, bhuiyan2023newscomp,petridis2023anglekindling, park2020rtag, bhuiyan2018feedreflect}.
For example, in supporting comprehension of news reading, \citet{chen2023marvista} explored a human-AI collaboration tool Marvista that employs a natural language processing model like abstractive summarization to provide text-specific support for reading articles in online news.
Through the main user study, they demonstrated the effectiveness of Marvista in enhancing users' comprehension of news content \cite{chen2023marvista}.
In addition, \citet{bhuiyan2023newscomp} developed and deployed Newscomp which allows users to annotate the similarities and differences between two different news sources, promoting diversity in reading through comparative annotation.
Nevertheless, they focus on comprehending the news, without incorporating further critical elements in the assisting process.
\kangyu{
\citet{feng2023topicbubbler} introduced TopicBubbler, a visual analytics system that facilitates the integration of fragmented information extracted across different levels and topics, supporting multi-level, fine-grained exploration of social media data. 
However, the system primarily focuses on aggregating large volumes of fragmented data, which may be less suited for we-media news, which may have limited information sources \cite{juneja2022human, wang2021journalistic}, and does not explicitly address how extracted information can be further leveraged to foster readers’ critical thinking.
}
As for critical news reading, \citet{Petridis2023} proposed AngleKindling, an interactive tool that utilizes the common sense reasoning capabilities of large language models to assist journalists in exploring angles for reporting on press releases.
To assist critical thinking in the identification of potential misinformation, \citet{park2020rtag} developed rTag View, a mobile news app user interface that distinguishes sections that need to be verified (e.g., allegation, assertion) with the confirmed truth, before making definitive factual judgments.
More, \citet{bhuiyan2018feedreflect} developed a browser extension, FeedReflect, that pushes users to pay more attention and use static reflective questions to assess the credibility of news on Twitter.
Through visual cues such as highlighting and blurring tweets, FeedReflect guides users to consciously assess which online content is credible \cite{bhuiyan2018feedreflect}.

In line with these tools, we design an interactive tool \Toolname{} that introduces similar features like classification and poses critical thinking questions. 
However, unlike these tools that focus on news content, \Toolname{} is a comment-powered assistant tool for supporting critical news reading in social media.

\cscwrr{
\subsection{Pre-trained Large Language Model and Multi-source Summarization}
The rapid development of pre-trained large language models (LLMs) has transformed natural language processing \cite{Min2023}. 
Trained on massive corpora, LLMs such as the GPT series \cite{radford2018improving} and more recent open models exhibit strong capabilities in language understanding, reasoning, and text generation across a wide range of downstream tasks \cite{zhang2023survey}.
Instead of task-specific fine-tuning, many systems now leverage LLMs through zero-shot or few-shot prompting, which requires less computation, enables rapid iteration, and removes the need for dedicated fine-tuning datasets \cite{liu2023pre}. 
This “prompt-based” use of LLMs has already been widely adopted in HCI tools that embed NLP components into interactive workflows \cite{scao2021many,petridis2023anglekindling}.
Consistent with other HCI systems that rely on prompting rather than fine-tuning, we adopt the pre-trained LLM Mistral-7B as the backend engine and iteratively develop prompt templates for the different text processing tasks in \Toolname{}’s workflow \cite{zamfirescu2023johnny}. 

In addition, recent work on multi-source summarization highlights both the promise and limits of LLMs in aggregating information from multiple heterogeneous sources. 
On the one hand, LLMs can produce fluent, high-coverage summaries and handle complex aggregation and reasoning over multiple documents or perspectives \cite{hosking2023attributable, huang2024embrace, kurisinkel2023llm,  nandy2025language}. 
However, on the other hand, these models may also bring some new problems, like hallucinating unsupported content, struggling with faithfully attributing statements to specific sources, and under-representing minority or less salient viewpoints \cite{belem2025single, zhang2024fair, deas2025summarization}. 
To mitigate these challenges, previous researchers have proposed some effective methods, like \emph{attributable} or evidence-grounded summarization pipelines that explicitly link summary statements back to supporting documents or segments \cite{hosking2023attributable, nandy2025language}, benchmarks for representing and evaluating \emph{diverse perspectives} and fairness across social groups or stances \cite{zhang2024fair, deas2025summarization, huang2024embrace},
and mitigation strategies building on analyses of hallucination types and simple post-processing heuristic methods \cite{belem2025single}.

Our work is aligned with this line of multi-source summarization research in two ways. 
First, \Toolname{} treats news and user comments as a multi-source input space: main points and critical hints are generated by aggregating content from the news together with a set of comments. 
Second, motivated by the literature on attributable and perspective-aware summarization, our design explicitly links generated main points back to their underlying comments, and aims to surface multiple angles rather than a single “consensus” view. 
While Mistral-7B is no longer the most state-of-the-art model now, our workflow is model-agnostic: stronger or more specialized LLMs for multi-source, attributable, and fair summarization could be plugged into the same pipeline to further improve summary quality, faithfulness, and coverage of diverse perspectives.
}

\section{Formative Study}

To support the critical news reading process and the need for guiding formative study design, we first constructed a 3R reading framework based on the literature. We then conducted online one-on-one meetings to engage participants in real reading experiences, followed by semi-structured interviews to identify the challenges and support needs users faced along the way.

\subsection{3R Critical Reading Framework} \label{sec:3r}
We modified the SQ3R ("Survey, Question, Read, Recite, Review") process, commonly used in reading skills training courses \cite{dharma2013implementation}, which allows readers to conduct a preliminary observation of the text (Survey), compile a list of questions (Question), read the whole text (Read), remember information  (Recite), and review the text to gain more information (Review) \cite{brown2015teaching}.
For example, \citet{dharma2013implementation} showed that applying SQ3R to news reading teaching improved students' reading ability. 
Specifically, we adapted SQ3R and proposed a \textbf{3R} reading framework: \textbf{R}ead content) Read the news content; \textbf{R}ead comment) Read the comments; \textbf{R}eflect) Reflect on the news with critical thinking questions.
Inspired by previous studies on users' reading preference \cite{machete2020use} and our specific reading scenario (\eg without well-framed news structure), we removed the Survey and Recite to reduce the user's workload.
Considering that users should read news content and read comments, we set two reading stages, \ie \cscwfr{\textbf{R}ead content and \textbf{R}ead comment}.
Finally, we merged the Question and Review stages into a Reflect stage where users should raise critical thinking thoughts. 
Unlike the original Question stage in SQ3R which requires more background knowledge and focuses more on text comprehension, we chose to put the Question stage after the Read stages and focused on critical thinking questions for facilitating the review process. 
Besides, since we aim at supporting reading tasks, we did not require readers to raise or write down their own questions after the Read stages as did in previous teaching \cite{dharma2013implementation} or training tasks \cite{yuan2023critrainer}. 


\subsection{Reading \& Interview}
\subsubsection{Participant}
We developed the recruitment questionnaire on Microsoft Forms, posted it to the social network in a local university, and invited respondents to complete it online. 
Among \ky{31} respondents, we invited \ky{12} participants (\ky{six} males, \ky{six} females, age: mean = \ky{21.42}, SD = \ky{1.51}) who covers reading comments very seldom (3/12) to very often (2/12) \cscwfr{, from different disciplines, such as artificial intelligence (5), software engineering (2), education (2), international relations (2) and philosophy (1)}. 
In addition, in a single-choice question with explanations of CNR, two participants reported that they ``did not know CNR'', seven participants ``knew it but seldom conducted it in practice, and three participant ``often conducted CNR''. 
All participants speak English as their second language, and they receive the CET-6 certificate, a national test indicating students’ English level of non-English major postgraduates in China, meaning that are qualified in reading and writing in English in our later task. 

\subsubsection{Reading Experience}
In order to obtain the real-life difficulties and needs that users encounter when reading news critically on social media, we first invited participants to conduct a critical reading of social media news in combination with the comments.
We chose the news written in English and posted by personal accounts on Twitter for the reading task, following previous work that studied news reading \cite{spinde2023twitter, verbeke2017critical, kunnath2019developing}.
Participants need to read two news with over 500 comments (noted as hot-topic news) and two news with factual errors (noted as fake news).
We chose these two types of news because hot-topic news can provide ample comments while discerning fake news requires a high level of critical thinking \cite{badawy2019falls, aimeur2023fake}.
For hot-topic news, two authors browsed and discussed the news on Twitter by the time of January 2024 and selected one article about court rulings \footnote{\url{https://x.com/LasVegasLocally/status/1742670392849060162}} (with 6k+ comments by that time) and the other article about policy \footnote{\url{https://x.com/JoJoFromJerz/status/1742388240697413784}} (700+ comments). 
For the two fake news, two authors repeated the process and chose one article about public safety \footnote{\url{https://x.com/pplatesrgrate/status/1729601773432668491}} (700+ comments) and the other article about business \footnote{\url{https://x.com/DrLoupis/status/1734728628884120019}} (3.2k+ comments), \ky{which are checked false \footnote{\url{https://www.reuters.com/fact-check/video-shows-suspected-arson-northern-england-not-an-electric-bus-fire-2023-12-06/}} \footnote{\url{https://www.reuters.com/fact-check/video-does-not-show-thrown-out-zara-clothes-amid-calls-boycott-2023-12-15/}} by Reuters, a well-known news agency.}
We did not examine or remove any comment of these articles at this stage as we would like the participants to flag the helpful or unhelpful comments for critical news reading. 
For each of the four articles, participants were told that they could read the article and its comment in the Twitter (or X) web interface for about seven minutes, followed by three minutes for verbally describing their understanding and opinions of the news. 

\subsubsection{Semi-structured Interview}
Semi-structured interviews were conducted with participants after they completed their reading experience. 
First, we asked participants about the difficulties and needs they encountered in reading news content and comments critically.
Then we asked participants whether the comments promoted their critical reading process.
After that, we specifically asked participants which comments they considered useful or peripheral for critical thinking and prompted them to provide reasons for their classifications.
\cscwrr{
For the two fact-checked misinformation articles, we also invited participants to explain how they judged the truthfulness of the content and why, using these discussions as probes to help them reconstruct and reflect on their own critical reading process rather than as a formal fake news detection task.
}
Finally, we introduced our proposed 3R reading framework and asked what kind of assistance participants needed in each stage.

\subsection{Findings}
All 12 participants agreed that some comments under the news were helpful for conducting critical news reading. 
They actively flagged the useful or peripheral comments for critical news reading during the semi-structured interview. 
In the following subsubsections, we first summarize the challenges participants encountered in critical news reading, followed by the classification of comments for critical reading and the perceptions of proposed expected features.

\subsubsection{Challenges of Critical News Reading}

\kangyu{
\textbf{C1: Feel uneasy to capture the news event-related information.}
Readers need not only to understand the ideas conveyed by the news itself, but also to understand the covered events.}
During the interview, 11 interviewees said that the news itself provided a relatively little description of the event (\eg time, location, causes, results, etc), and it was difficult for them to mine information related to the event from comments. 
``\textit{I feel there is very little description from the publisher about the news, and I have to constantly piece together information from the comments to understand what happened}'' (FS 1 male, age: 22).

\kangyu{
\textbf{C2: Feel uneasy to access meaningful information from news comments.}
10 interviewees said when reading the comments, they found the volume of information was substantial, but many comments lacked meaningful content, which interfered with the information gathering process and imposed a significant cognitive burden, made forming a comprehensive understanding quite challenging.}
One participant said that she tends to only read the most popular comments in daily life and misses many valuable comments. 
\textit{``I find it more exhausting than my usual reading routine. Normally, I just skim through some top comments, but it may not give me a comprehensive understanding''} (FS 3, female, age: 21).

\textbf{C3: Lack of skills in raising critical thoughts.}
\ky{11} interviewees stated that when reading the news, they lacked the ability to analyze content from multiple perspectives, \eg what keywords can serve as entry points, and how to criticize them.
``\textit{I find critical reading challenging and I usually only focus on comprehension. I may not come up with many ideas on my own, so I rely on comments to see how others think}'' (FS 2, male, age: 22).
In addition, participants generally indicated a lack of consideration for the rationality of their own viewpoints.
``\textit{Some of my ideas may come from inspiration or from other people's comments, but I seem to rarely consciously consider whether they are reasonable}'' (FS 6, female, age: 20).

\subsubsection{Comment classification for critical news reading} \label{sec:comments_types}
We utilize a thematic analysis \cite{braun2006using} to synthesize users' responses during the interview to what types of comments are useful/peripheral during the critical reading process.
First, we transcribed a total of four hours of interview recordings with 12 participants into textual interview records.
Then, two researchers independently analyzed and coded descriptions of useful and peripheral comments in the interview records.
After that, the two researchers got together to compare the contents and categories they had coded.
They merged similar categories based on their discussion and refined the descriptions and definitions of the categories.
Finally, they discussed inter-code connections and clustered related codes, resulting in high-level themes. 
In total, we collected 60 useful comments and 51 peripheral comments along with the reasons that helped us build up the taxonomy.
These results validate the assumption from the participants that many comments can be useful for critical news reading, while many others could be distracting.
Table \ref{classification} presents the final comment categories that are useful or peripheral in the critical reading process.
\begin{table}[htbp]
    \caption{Themes and comment categories are developed by synthesizing user descriptions of both constructive and less constructive comments. For each category, a definition and a representative example were provided.}
    \centering
        \begin{tabular}{m{2cm} m{2.2cm} m{5.1cm} m{3.2cm}}
            \toprule
            \textbf{Theme} 
                & \textbf{Category} 
                & \textbf{Definition} 
                & \textbf{Example} \\
            \midrule
            \multirow{2}{=}{\textbf{Information Enrichment}}  
                & \textbf{Contextuali\-zation}   
                    & Comments that rephrase, summarize, or elaborate on the original post to provide additional context or make the content clearer for readers. 
                    & ``Context/Details: In a shocking turn of events ...'' 
                \\ \cmidrule{2-4}
                & \textbf{External \newline Information} 
                    & Additional context or details from outside sources that enhance understanding of the event
                    & ``Iceland official source talk about that ...'' 
                \\ 
            \midrule
            \multirow{3}{=}{\textbf{Personal Engagement}} 
                & \textbf{Analysis}             
                    & Interpretation, reasoning, or critical assessment of the news content, often expressing personal viewpoints or arguments.
                    & ``The source article can not be corroborated from official channels'' 
                \\ \cmidrule{2-4}
                & \textbf{Association}          
                    & Personal experiences or references to related events that establish a connection to the news topic.
                    & ``I am from Iceland, it's true ...'' 
                \\ \cmidrule{2-4}
                & \textbf{Attitude}             
                    & Expression of a positive, neutral, or negative stance toward the news or individuals involved.
                    & ``Seems like she made the right call'' 
                \\
            \midrule
            \multirow{2}{=}{\textbf{Critical Reflection}}       
                & \textbf{Skepticism}            
                    & Questioning the accuracy, truthfulness, or intent of the publisher's content or the event itself.
                    & ``Why are you lying and saying this is London?'' 
                \\ \cmidrule{2-4}
                & \textbf{Provocation}           
                    & Introduction of new perspectives, prompts, or critical questions that encourage further thought or investigation.
                    & ``Stop posting AI videos''
                \\
            \midrule
            \multirow{4}{=}{\textbf{Peripheral Content}}
                & \textbf{Entertainment}        
                    & Humorous, sarcastic, or mocking remarks targeting individuals or events.
                    & ``Was there a trampoline in the court room?'' 
                \\ \cmidrule{2-4}
                & \textbf{Polarization}         
                    &  Strongly emotional or divisive comments that amplify disagreement or conflict.
                    & ``Idiota, you silly'' 
                \\ \cmidrule{2-4}
                & \textbf{Advertisement}        
                    & Promotional comments endorsing products, services, or platforms, often unrelated to the news topic.
                    & ``Google my book ...'' 
                \\ \cmidrule{2-4}
                & \textbf{Nonsense}             
                    & Comments that lack meaningful content or relevance to the news topic, such as references to other users or accounts.
                    & ``@xxx(account)'' 
                \\
            \bottomrule
        \end{tabular}
    \label{classification}
\end{table}
\subsubsection{Proposed Expected Features}
In the interviews, participants actively proposed the features they expected for a critical news reading support tool. 
In the \cscwfr{Read content} stage, participants suggested that if the system provides users with a summary (\textit{N}umber of participants mentioned = 11) and indices of the news-related information (N = 9), it would reduce their burden of searching for information. ``I hope this tool can automatically summarize information about the news from comments, which can help me understand the news content more quickly'' (FS 4, female, age: 21). 
In the \cscwfr{Read comment} stage, 11 participants expressed a desire to choose useful comments for critical reading, avoiding the distraction of numerous meaningless
comments.
Meanwhile, 10 participants also would like to see an overall distribution of attitudes among others.
``\textit{If the tool could help filter out these unhelpful comments, I could devote more time to analysis rather than searching for useful information}'' (FS 1, male, age: 22).
Lastly, in the Reflect stage, all participants expressed a high level of appreciation if the tool could provide reference angles and questions that inspire their critical thoughts.
``\textit{I strongly desire the tool to provide angles for reflection and critical thinking questions, helping me grasp the focal points of consideration quickly}'' (FS 5, female, age: 22).

\subsection{Design Requirements}
\label{sec:design_requirements}

Based on the findings and existing literature on users' reading behaviors and critical reading, we derive the following design requirements (DR) for a critical news reading support tool.

\textbf{DR1: To support reading of the news (\cscwfr{Read content}), the tool should summarize news-related information from the comments, and identify the sources of corresponding comments. } 
Summarizing the main article along with comments containing \textbf{Information Enrichment} (\autoref{classification}) can complement news details about the event's cause, process, or impact \cite{diakopoulos2011towards, williams2021effects} and help with reducing retrieving cognitive burden, which fitting readers' cognitive heuristic reading tendencies \cite{machete2020use}.
In addition, providing the summary of news-related information can help reduce the cognitive burden of information gathering, which is consistent with readers' cognitive heuristic reading tendencies \cite{machete2020use}.

\textbf{DR2: To support reading of the comments (\cscwfr{Read comment}), the tool should provide the features of filtering useful comments, and provide the proportion of different types and attitudes of comments.}
\yuan{Comments under the themes \textbf{Personal Engagement} and \textbf{Critical Reflection} (\autoref{classification}) can inspire critical thinking \cite{diakopoulos2011towards,kim2021improving}.}
Peripheral comments (such as Entertainment or Polarization) not only fail to support readers' critical thinking but also diminish their desire to read \cite{park2016supporting, diakopoulos2011towards}.
By filtering out the peripheral comments we identified, readers can concentrate their time on reading the useful comments, leading to a better reading experience \cite{kim2021improving}.
Additionally, as participants anticipated, the tool should provide statistical proportions for each type and attitude of comments, helping them to intuitively understand the distribution of comments.
 
\textbf{DR3: To support reflection on the information (Reflect), the tool should provide multifaceted keywords and thought-provoking critical thinking questions}.
Keywords can help readers quickly identify points of analysis in news while corresponding critical thinking questions can serve as good starting points to stimulate readers' thinking, as is the case with many other critical thinking tools \cite{yuan2023critrainer, petridis2023anglekindling, richards2023readerquizzer}.
%
%

\section{System}
\label{sec:system}
\begin{figure}[htbp]
\centering
\includegraphics[scale=0.35]{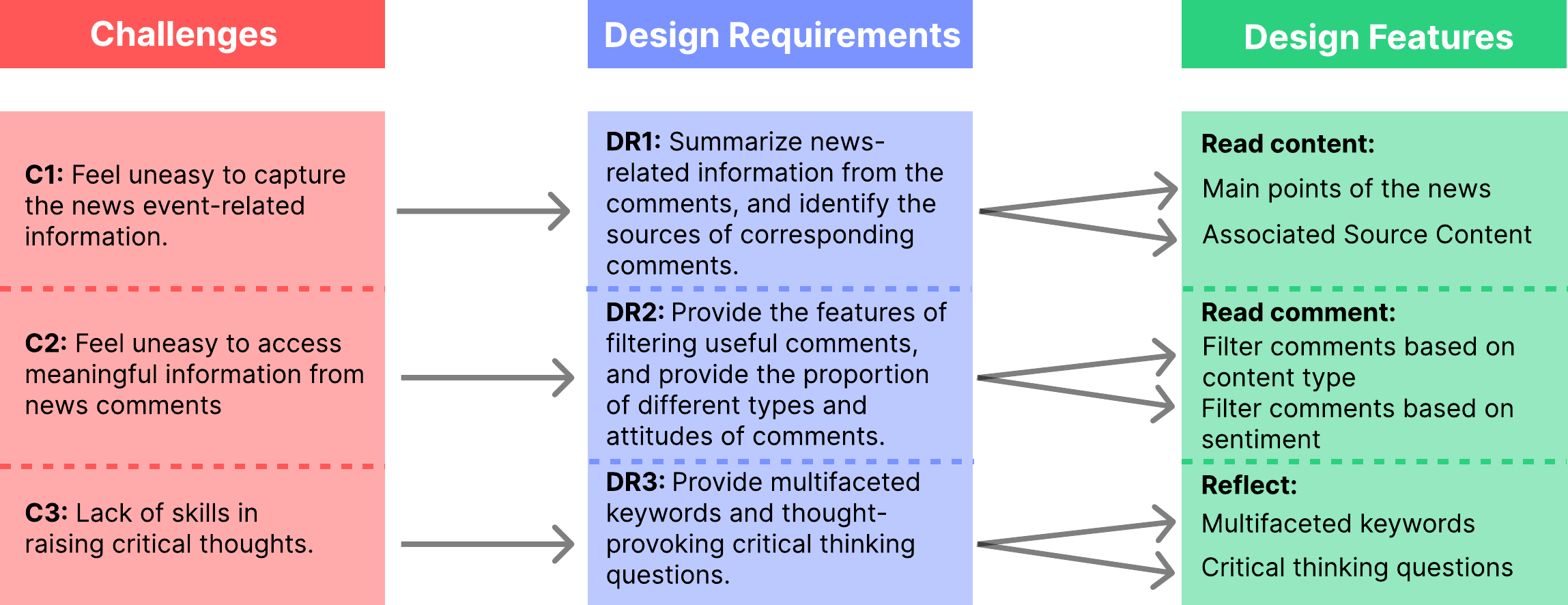}
\caption{Design pipeline of \Toolname{}}
\label{Design pipeline}
\end{figure}

Based on the design requirements for the comment-powered critical news reading support tool, we developed \Toolname{}, which provides a summary of the main points for the news combined with event-related comments in the \cscwfr{Read content} stage (DR1), supports users to filter comments that could be useful for critical thinking in the \cscwfr{Read comment} stage (DR2), and offers heuristic machine-generated keywords and corresponding critical thinking questions in the Reflect stage (DR3). 
\kangyu{Figure \ref{Design pipeline} shows the design pipleline of \Toolname{}.}
We choose to develop and deploy \Toolname{} locally to avoid potential interference from other extraneous factors, such as extra built-in features (\eg Navigation sidebar and Search box).
Our interface design layout is inspired by the popular social media platform Twitter \cite{karami2020twitter}, but it can be generalized and customized to other platforms such as Facebook, and Weibo.
The \Toolname{} is implemented in Vue.js and connects to the backend Python flask server that processes the news content, comments, and user interaction.
In this section, we provide a detailed description of the interface design and backend models in \Toolname{}. 

\subsection{User interaction with \Toolname{}}
\begin{figure}[htbp]
\centering
\includegraphics[scale=0.09]{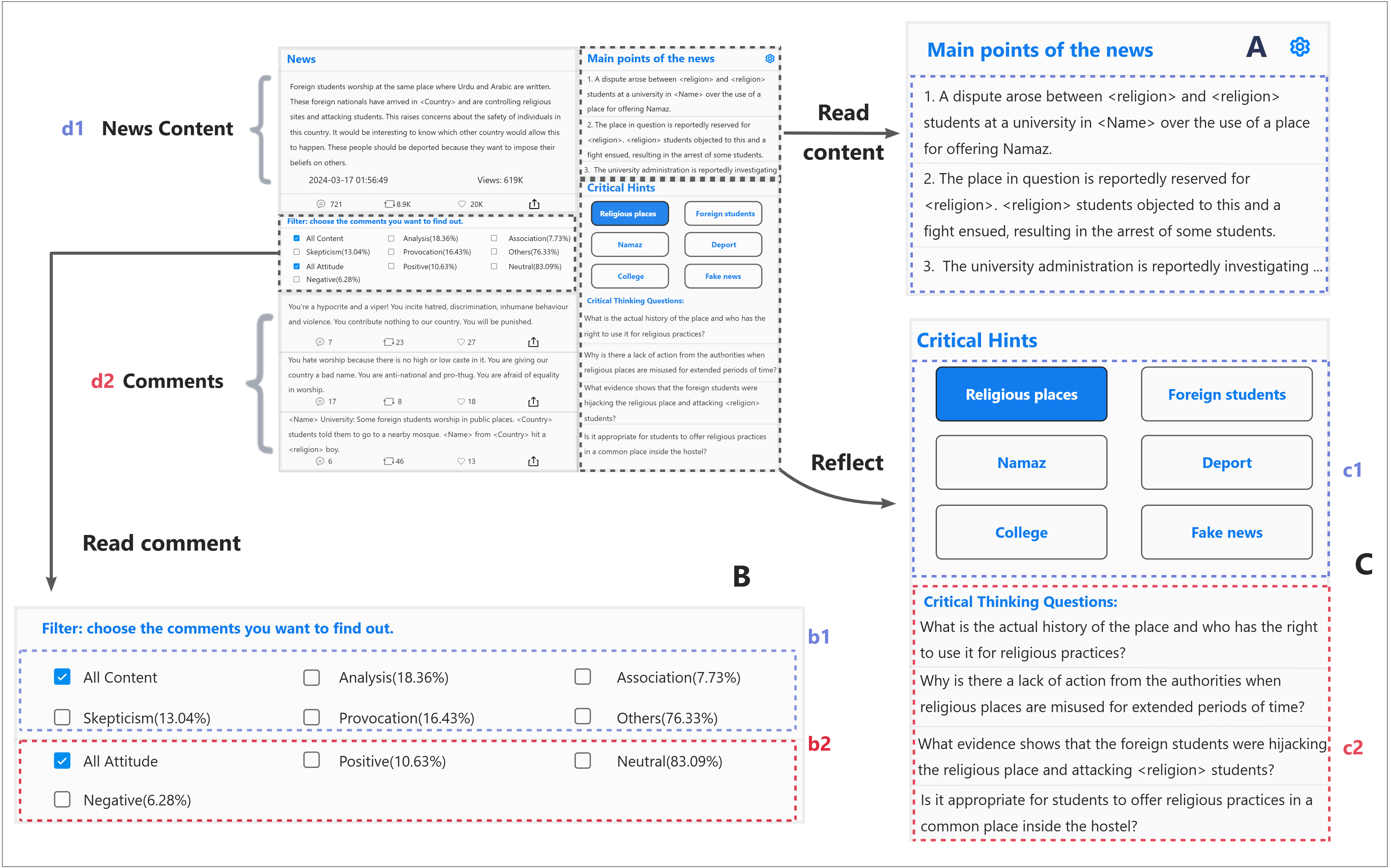}
\caption{Interface of \Toolname{} in the 3R reading process. Users can read the news content (d1) and comments (d2). In \cscwfr{Read content} stage, users can read the main points (A) and click interesting points to get related comments (d2). In \cscwfr{Read comment} stage, users can filter the types of comments they want to browse (B) based on the content (b1) or sentiment (b2). In Reflect stage, users can raise critical thinking thoughts with the support of keywords (c1) and corresponding critical thinking questions (c2). For privacy concern, we rephrase the news content and comments, and replace sensitive information using templates (\eg <Name>, <Country>, and <religion>).}
\label{interface}
\end{figure}

Figure \ref{interface} illustrates the interface design of \Toolname{}.
In addition to the content and comments of news that appears like those in other social media interfaces, 
\Toolname{} provides unique features tailored to the 3R reading process, which includes the key points of the news (\cscwfr{Read content}), comments filter (\cscwfr{Read comment}), and keywords with corresponding critical thinking questions (Reflect).
We introduce the three stages respectively.

\textbf{\cscwfr{Read content}:}
In this stage, users should have a comprehensive and correct understanding of the main content of the news, such as the cause, process, and outcome.
Users can scroll up and down to view the bullet key points of the news (Figure \ref{interface}A), which are summarized based on the news content and the comments classified into \textit{Contextualization} or \textit{External Information} (Table \ref{classification}).
Users can click any key point of the news, which shows the corresponding comments that contribute information to this point in the comment section. 
Users can click the selected key point again to deselect it. 

\textbf{\cscwfr{Read comment}:}
In this stage, users should read the comments to learn others' opinions and perspectives. 
\Toolname{} provides a comment filter (Figure \ref{interface} B) for this purpose.
The upper of the comment filter (b1) allows users to filter comments based on the type of content, \ie \textit{Analysis}, \textit{Association}, \textit{Skepticism}, and  \textit{Provocation} (Table \ref{classification}).
The \textit{All Content} option will list all the comments in the comment section, 
while
\textit{Others} represents a collection of comments that could be peripheral for critical thinking, \ie those are classified as \textit{Entertainment}, \textit{Polarization}, \textit{Advertisement}, and \textit{Nonsense} (Table \ref{classification}). 
The lower part of the comment filter (b2) allows users to select comments based on their attitudes toward comments, \ie \textit{positive}, \textit{neutral}, and \textit{negative}. 
If users select multiple options in the content filter or the attitude filter, \Toolname{} will display the union set of the filtered comments. 

\textbf{Reflect:}
In this part, users should engage in critical thinking about the news or comments.
\Toolname{} provides users with \cscwrr{\textbf{keywords (c1)}} and corresponding \cscwrr{\textbf{critical thinking questions (c2)}} for reference, which are extracted from and generated based on the comments labeled \textit{Skepticism} and \textit{Provocation} (Table \ref{classification}). 
\cscwrr{Users can click to toggle between \textbf{keywords in c1} to switch the relevant \textbf{critical thinking questions in c2}.}

\begin{figure}[htbp]
\centering
\includegraphics[scale=0.7]{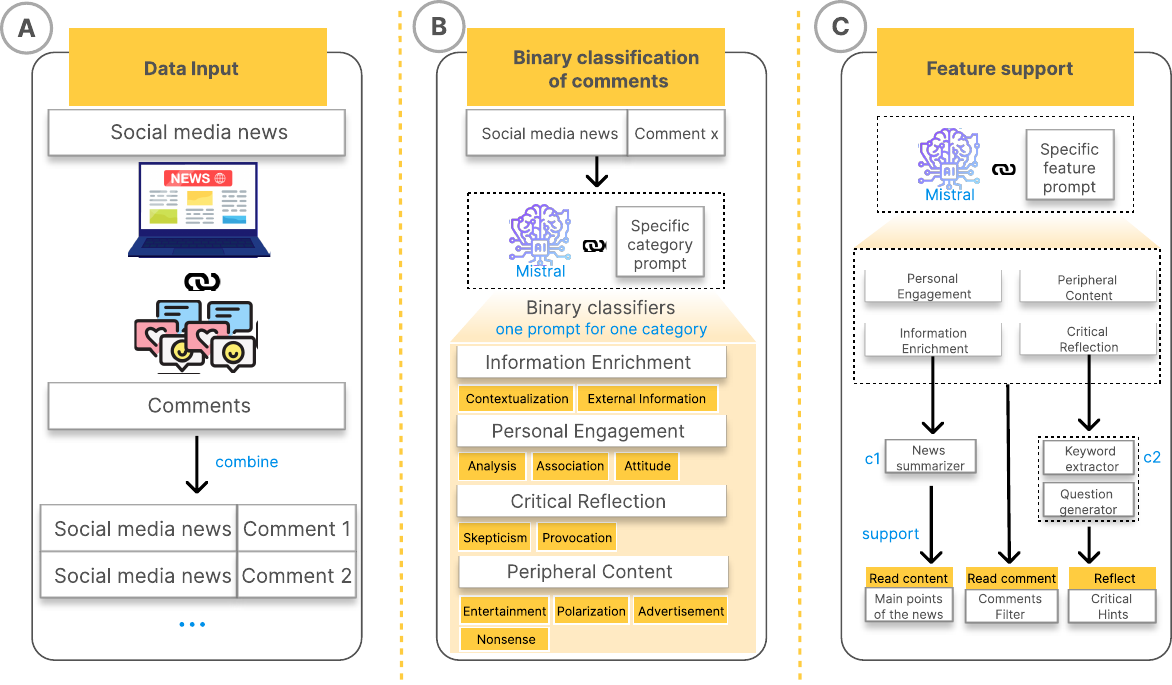}
\caption{NLP pipeline. News content and its first-level comments (A) are concatenated into a news-comment pair and input into the backend consisting of the Binary Classifiers (B), Summarizer (c1), and Keywords Extractor \& Question Generator (c2), which are supported by Mistral 7B. The Binary Classifiers perform classification for the comment on each category in the table \ref{classification} to support \cscwfr{Read comment} stage. The information-related comments are fed into Summarizer for main points generation and relevance identification to support \cscwfr{Read content} stage. The inspiring comments are fed into Keywords Extractor \& Question Generator for providing keywords and critical thinking questions to support Reflect stage.}
\label{backend}
\end{figure}

\subsection{Backend}
Figure \ref{backend} illustrates the natural language processing pipeline that enables user's interaction with \Toolname{}. 
For each first-level comment of news, \Toolname{} first concatenates it with the content of the news to form a joint entity of news-comment pair (Figure \ref{backend} A).
We choose the first-level comments because they have a direct relevance to the news content.
Then, our binary classifiers take each news-comment pair as input and predict what the content and sentiment of each comment is about (Figure \ref{backend} B).
Next, the comments predicted as \textit{Contextualization} or \textit{External Information} (Table \ref{classification}), together with the news content, are sent to the summarizer (Figure \ref{backend} c1) that generates a list of comment-powered key points of the news to support users in the \cscwfr{Read content} stage. 
For all comments, the tags assigned by our binary classifiers empower the filters in the \cscwfr{Read comment} stage. 
Lastly, the comments inferred as \textit{Skepticism} or \textit{Provocation} (Table \ref{classification}) are sent to the keyword extractor and question generator (Figure \ref{backend} c2) that outputs thought-provoking keywords and corresponding critical thinking questions to support users in the Reflect stage. 
Figure \ref{Model_building_process} shows the process of how we build up our backend support models.

\subsubsection{Binary classifiers of news comments}
\label{binaryclassifiers}
\textbf{Data Collection.}
\cscwrr{Two researchers collected and applied the taxonomy to 240 comments from 16 news articles on different topics (\eg policy, health, education, gender, violence, science, technology). This data is used to develop LLM-based binary (yes/no) classifiers, one for each comment category, which take a news content–comment pair as input and output a yes/no label indicating whether the comment belongs to the category.}
We add the corresponding news content to each comment because the classification of comments needs to be reflected in the corresponding context.
First, the two researchers browsed Twitter independently, and collected and coded 20 comments with their news, making sure that each label in each category was covered.
We use Gwet's AC1 to measure the consistency of independent coding instead of the other method (\ie Cohen's $\kappa$) because it is more applicable and stable when certain metrics are sparse (\eg Contextualization, External information). 
The statistical result shows that they achieved a high level of agreement (Gwet's AC1 $\geq$ 86.6\%).
They discussed and recorded the comments which shows inconsistency and then reached a consensus.
They repeated the process, achieved substantial agreement on these categories (Gwet's AC1 $\geq$ 93.9\%), and independently collected and coded an extra 100 comments.
Finally, we obtained a total of 240 coded comments from 16 news (\eg Policy, Health), with more than 20 examples for each label in each category.
For each label in each category, we selected 20 examples from all coded ones, accompanied by their news, ultimately composing the binary classification dataset.

\begin{figure}[htbp]
\centering
\includegraphics[scale=0.6]{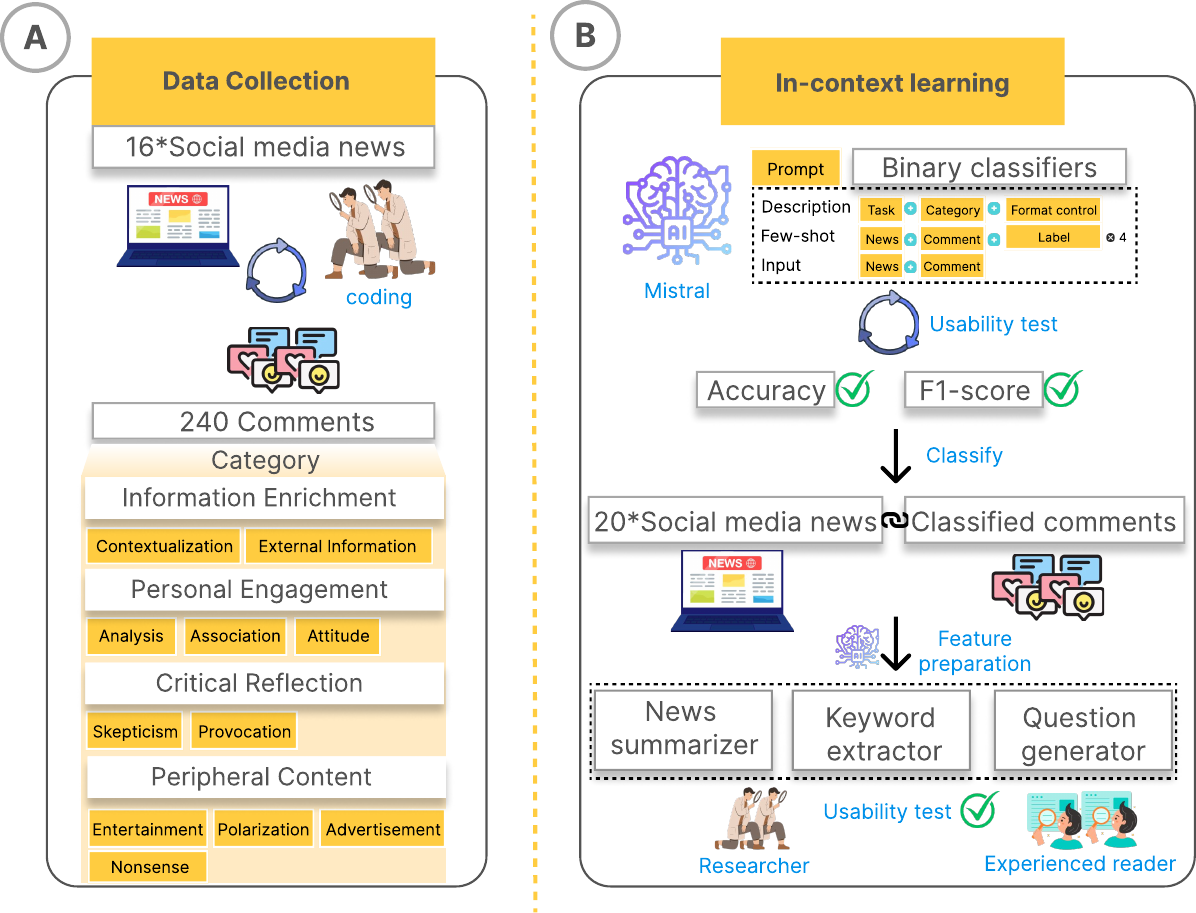}
\caption{Process of building up models. Two researchers first looked through 16 social media news and coded 240 comments. The labeled comments are put into in-context learning for building up the binary classifiers based on Mistral 7B. Next, the binary classifiers which passed the usability test are used to classify the comments from 20 social media news. These social media news and classified different types of comments are combined for building up the News summarizer, Keyword extractor, and question generator, which were validated by two researchers and two experienced social media news readers.}
\label{Model_building_process}
\end{figure}

\textbf{Prompt learning for binary classifiers.}
We choose to use a large language model, Mistral 7B \cite{jiang2023mistral}, as our backend model.
Based on the traditional iterative prototyping process for HCI from \cite{zamfirescu2023johnny}, we prepared a separate prompt for each category, but with a similar format shown in figure \ref{Model_building_process}.
We first designed the basic prompts based on existing prompt engineering documents \cite{OpenaiPromptEngineer}.
We included the news content as part of the prompt to provide contextual context.
For the binary classification, we also used a few-shot for better performance, \ie putting two extra examples for each label in each prompt.
Then, we tested the prompt using the prepared binary classification dataset.
The researchers reviewed the misclassified samples and modified the prompt accordingly.
We repeated this process until the prompt achieved a usable performance, with at least $0.75$ in accuracy and $0.71$ in F1-score for each category.

\subsubsection{News Summarizer}
\label{summarizer}
\textbf{Data collection.}
To measure the performance of the prompts on main points generation and relevance identification, two researchers browsed Twitter and then collected 20 news with thousands of comments and thinking-valued topics like Policy and Health.
By employing a Twitter crawler, we scraped the news content as well as all publicly available comments.
For the comments on each news, we first classified them using the mentioned binary classifiers in \ref{binaryclassifiers}.
Then, for each news, we selected the comments labeled as \textit{Contextualization} or \textit{External information} for further processing with the news summarizer.

\textbf{Prompt learning for news summarizer.}
The news summarizer should first generate the main points of the news based on the news content and information-related comments.
Then, for each key point, the news summarizer should find relevant comments associated with each main point.
We randomly selected 4 news from the collected news for iteratively optimizing prompts during prompt learning.
We first input the news content, information-related comments, and the main points generation prompt into the LLM to obtain the main points of the news. 
Then, we input the information-related comments, the main points, and the relevance generation prompt into the LLM to obtain relevance between the points and comments.
Two researchers evaluated the quality of the generated key points and relevance adapted from \cite{kovacs2014smart}'s three metrics that address summary:
\textit{Readability:} To what extent is the generated summary understandable? 
\textit{Completeness:} To what extent does the generated summary cover the original content? 
\textit{Relevance:} To what extent is the generated citation serial number corresponding to the original text relevant to the main points?
We repeated this process until both researchers agreed that the generated content supported the use of the tool in these metrics.

\textbf{Performance evaluation.}
We processed the remaining 16 news using the obtained main point generation prompt and relevance identification prompt.
Then, we invited two social media news readers unrelated to this study to evaluate the machine-generated content, who self-reported that they had a lot of reading experience in social media.
They are required to score machine-generated main points and relevance with the three metrics mentioned above with a five-point Likert Scale. 
Additionally, they are required to rate their perceived truthfulness of the generated key points on a five-point Likert Scale (1/5 - strongly disagree/agree), using two metrics adapted from \cite{soprano2021many}, \ie
\textit{Precision} (``the generated key points are precise enough rather than vague'') and
\textit{Trustworthiness} (``the generated key points are generally trustworthy compared with original news and comments'').
In either evaluation task, before independently rating all generated points, the raters first familiarized themselves with several data samples and reached an agreement on the rating scheme adapted from the suggestions of GPT-4 (Appendix \ref{rubrics}).
We ultimately averaged their scores and got the following results:
\textit{Readability}: $M=3.800$, $SD = 0.870$; \textit{Completeness}: $M = 3.625$, $SD = 0.881$; \textit{Relevance}: $M = 3.600$, $SD = 0.682$; \textit{Precision}: $M = 3.675$, $SD = 0.859$; \textit{Trustworthiness}: $M = 3.750$, $SD = 0.899$. 
According to the rating schemes (Appendix \ref{rubrics}), these scores indicated that the generated summary is relatively understandable, adequately covers the original content, is reasonably relevant, offers fairly precise key points, and is generally trustworthy.



\subsubsection{Keyword Extractor and Question Generator}
\textbf{Data collection.}
To measure the performance of the prompts on keyword generation and question generation, we adopted 20 news the same as \ref{summarizer}.
Differently, we combined news content and comments labeled as \textit{Skepticism} or \textit{Provocation} (\autoref{classification}) for keyword and question generations, because these types of comments often disagree with the article's message and were likely to affect the readers' opinions about the news events \cite{houston2011influence, lee2010others, walther2010influence}.

\textbf{Prompt learning for keyword extractor and question generator.}
The keyword extractor should extract keywords that promote critical thinking based on the input (the news content and inspiring comments).
Then, the question generator should generate relevant critical thinking questions for each keyword.
Similar to \ref{summarizer}, we input the news content, inspiring comments, and keywords generation prompt into the LLM for extracting keywords.
Then, we input the news content, comments, generated keywords, and the question generation prompt into the LLM for obtaining the critical thinking questions.
Similarly, two researchers evaluated the quality of the generated keywords and critical thinking questions from \cite{le2015evaluation}'s three metrics that address critical thinking questions:
\textit{Relevance:} To what extent are the generated keywords related to the original content? 
\textit{Accessibility:} To what extent are the generated critical thinking questions understandable? 
\textit{Usefulness:} To what extent do the generated critical thinking questions help with raising further personal critical thinking thoughts?
We repeated this process both researchers agreed that the generated content supported the use of the tool.

\textbf{Performance evaluation.}
We also processed the remaining 16 news using the obtained keywords generation prompt and question generation prompt.
Then, the invited two scorers evaluated machine-generated keywords and critical thinking questions from the three metrics mentioned above with a five-point Likert Scale (1/5 - very bad/good), also based on the consensus rating scheme (Appendix \ref{rubrics}). 
We ultimately averaged their scores, and each metric passed the usability test (Relevance: $M = 3.475$, $SD = 0.688$; Accessibility: $M = 3.875$, $SD = 0.623$; Usefulness: $M = 3.6$, $SD = 0.758$). 
\revise{
According to the rating schemes (Appendix \ref{rubrics}), these scores indicated that the extracted keywords are relatively related to the original content, and the generated questions are fairly understandable and relatively useful.
To demonstrate the enhancement brought by the inclusion of comments, 
we prompted the model to generate a set of keywords and questions based on similar prompts, without the comment-related portions. 
The same two raters are required to evaluate the outputs under the same rating schema (Appendix \ref{rubrics}). 
The results in the three metrics are Relevance: ($M = 3.13$, $SD = 1.04$ ); Accessibility: ( $M = 3.5$, $SD = 0.64$ ); Usefulness: ( $M = 3.05, SD = 0.99$ ). 
The results of Wilcoxon signed-rank tests \cite{woolson2007wilcoxon} 
suggest that the LLM is capable of generating more relevant ($W = 41.0$, $p < 0.05$), accessible ($W = 17.5$, $p < 0.05$), and useful ($W = 15.0$, $p < 0.01$) keywords and critical thinking questions when including comments.
}

\section{Experiment}

Our evaluation of \Toolname{} aims at answering the following research questions (RQs):

\textbf{RQ1:} 
How would \Toolname{} help the social media users with (i) comprehending news content \ky{and context}, (ii) raising critical thinking thoughts, and (iii) self-perception? 

\textbf{RQ2:} 
How would \Toolname{} affect users' (i) behaviors and (ii) perceived engagement and workload during the reading process?

\textbf{RQ3:} 
How would users perceive the use of \Toolname{} for assisting critical news reading on social media?

Specifically, we conducted a within-subjects study with 24 participants who often read comments under news (at least five times per ten articles) to answer the three RQs via statistical comparisons to a baseline tool that simulates traditional reading experience. 
We also conducted a qualitative study with eight participants who do not often read comments to gain their opinions to complement findings about RQ2 and RQ3. 
In this section, we first describe the design of the within-subjects study, followed by a subsection that describes the qualitative study.

\subsection{Baseline}
To evaluate the value of our proposed interactive features, we choose a baseline tool that is 
similar to \Toolname{} but without its unique features of news summarization (\autoref{interface}A), comment filters (B), and critical questions (C).
Within both the baseline tool and \Toolname{}, users can imitate the reading process on Twitter, including viewing the publisher's description of the news, others' comments, and multi-level comments.
We used an open-sourced API tweety \footnote{\url{https://github.com/mahrtayyab/tweety}} to crawl the comments following the order of their appearances under the news in the Twitter web pages. 
This means that the order of comments in Baseline and \Toolname{} is consistent with the order of comments in the original Twitter page at the time of data crawling.

Overall, both tools simulate experiences of how users can read news and check others' comments on Twitter in their daily lives. 
However, we do not include other interactive features like navigation sidebars, likes, and forwards that are not directly related to our focused reading process. 
For interactive icons like `Like' in the system, there is no real feature. Readers are only permitted to see the numbers.
\yuan{Table \ref{Comparison} shows a comparative analysis of common and unique features among CoNewsReader, baseline, and existing reading platforms.}

\begin{table*}[htbp]
\captionsetup{width = \textwidth}
\caption{The comparative analysis of common and unique features among \Toolname{}, baseline, and existing reading platforms. }
\resizebox{\textwidth}{!}{
\begin{tabular}{c>{\centering}m{.105\textwidth}>{\centering}m{.1\textwidth}>{\centering}m{.1\textwidth}>{\centering}m{.1\textwidth}>{\centering}m{.105\textwidth}>{\centering}m{.1\textwidth}>{\centering}m{.1\textwidth}>{\centering}m{.1\textwidth}>{\centering}m{.11\textwidth}c}
    \toprule
                & {Post Showcase} & {Comment Area} & {News Summary} & {Comment Filters} & {Critical Questions} & {Reply} & {Repost} & {Like} & {Book\-mark} & {Share} \\ \midrule
    \Toolname{} & \checkmark                        & \checkmark                       & \checkmark                             & \checkmark                          & \checkmark                             &                           &                            &                          &                              &                           \\
    Baseline    & \checkmark                        & \checkmark                       &                                        &                                     &                                        &                           &                            &                          &                              &                           \\
    Twitter     & \checkmark                        & \checkmark                       &                                        &                                     &                                        & \checkmark                & \checkmark                 & \checkmark               & \checkmark                   & \checkmark                \\
    Facebook    & \checkmark                        & \checkmark                       &                                        &                                     &                                        & \checkmark                &                            & \checkmark               &                              & \checkmark                \\
    Weibo       & \checkmark                        & \checkmark                       &                                        &                                     &                                        & \checkmark                & \checkmark                 & \checkmark               &                              & \checkmark                \\
    Instagram   & \checkmark                        & \checkmark                       &                                        &                                     &                                        & \checkmark                &                            & \checkmark               & \checkmark                   & \checkmark                \\
    Quara       & \checkmark                        & \checkmark                       &                                        &                                     &                                        & \checkmark                &                            & \checkmark               & \checkmark                   & \checkmark                \\ \bottomrule
\end{tabular}
}
\label{Comparison}
\end{table*}

\subsection{Participants}
We recruited \yuan{24} students, consisting of \yuan{14} undergraduates and \yuan{10} postgraduates, through an advertisement posted on a social network in a Chinese university. 
There are \yuan{11} males and \yuan{13} females \yuan{(age: $M = 22.13$; $SD = 1.56$)}, \cscwfr{covering different disciplines, \eg computer science, artificial intelligence, history, tourism management, languages, and philosophy}.
All participants speak English as a second language, and they have obtained the College English Test (level 6) certificate.
Participants generally maintained daily social media news reading experience \yuan{($M = 1.68$ hours a day; $SD = 1.28$)} and self-rated reading comments in a moderate to high frequency (eight participants with ``5-6'', six with ``7-8'' and ten with ``9-10'' in reading comments frequency per ten news).
We randomly assigned participants into the four different groups in our Latin Square described in the next subsection.

\subsection{Task and Procedure}

\begin{figure}[htbp]
\centering
\includegraphics[scale=0.3]{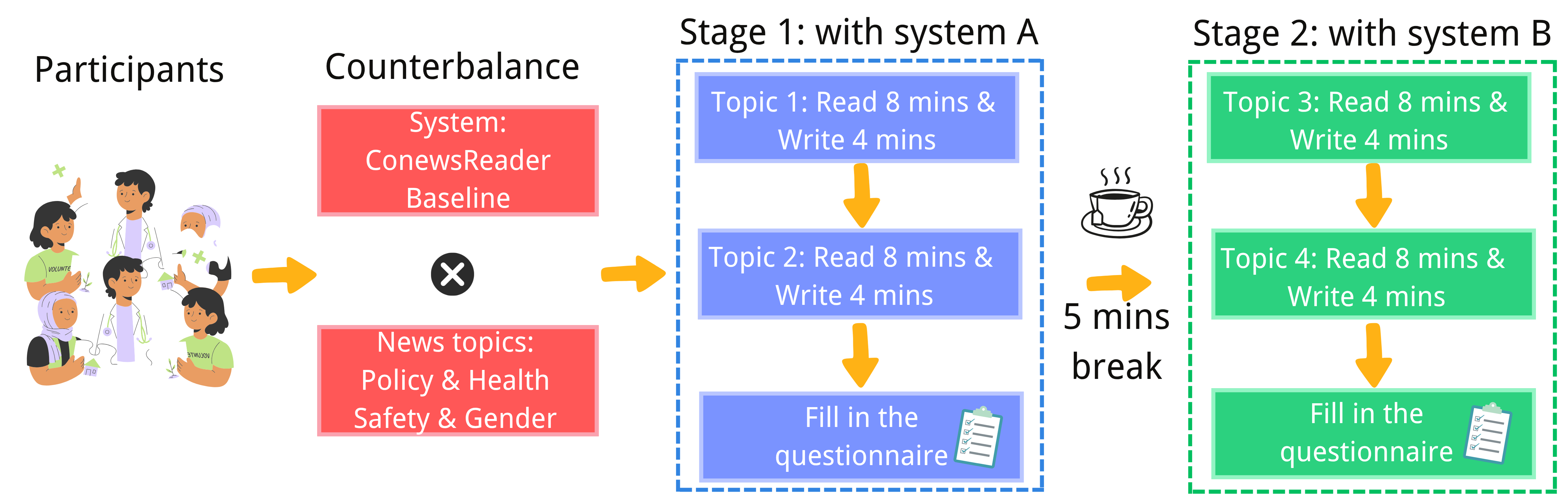}
\caption{Experiment design and procedure. 
}
\label{experiment}
\end{figure}

We conducted the experiment offline. 
Figure \ref{experiment} presents the procedure for each participant. 
Since \Toolname{} focuses on the scenario where users need to read critically with comments, we sample 20 news from Twitter that have thousands of comments. 
All of these news focus on topics worthy of consideration (\eg Policy, Health, Safety), which need critical thinking to avoid misleading (\ie receiving bad health advice).
After discussing and selecting based on the information load, we finally chose four posts that have roughly balanced information load and focused on the topics of policy, religion, gender, and health, respectively. 
With the Twitter crawler \textit{tweety}, we crawled all publicly available information (\eg post content and comments) as the reading material for user study. 
Due to the adoption of a within-subjects study design, we can disregard individual differences among participants under different reading support conditions.

We randomly divided the four articles into two groups and fixed them in two reading tasks. To eliminate the order effects of reading tasks and used tools, we adopted the experimental design of Latin Square:

\begin{itemize}
    \item Policy\&Gender (Baseline) - Health\&Religion (\Toolname{})
    \item Health\&Religion (Baseline) - Policy\&Gender (\Toolname{})
    \item Policy\&Gender (\Toolname{}) - Health\&Religion (Baseline)
    \item Health\&Religion (\Toolname{}) - Policy\&Gender (Baseline)
\end{itemize}

\textbf{Stage 1:}
We first introduced the background and procedure of the experiment to the participants, and all participants signed informed consent forms before the experiment.
Then, participants were randomly assigned to one of the groups in the Latin square. 
The task prompt is: ``You are going to conduct two critical news reading tasks. In each task, you need to read two news one by one using a given tool. After each task, we will ask you to write down your comprehension and critical thoughts about the news and rate your perceptions on the given tool''.
Subsequently, participants read two news (P\&G or H\&R) using either Baseline or \Toolname{} according to their assigned group.
For each news, participants had eight minutes to read and an extra four minutes to write down their comprehension and critical thinking thoughts about the news. 
During the writing period, they could not browse the news again.
After completing the reading and writing for the two news, participants were asked to fill out a questionnaire to collect their opinions on the tool they used.

\textbf{Stage 2:}
After a five-minute break, participants were required to use another tool to read the other two news, similar to Stage 1.
Subsequently, participants were also required to fill out a questionnaire to gather their opinions on the tool used in Stage 2.

\subsection{Measurement}
\textbf{RQ1: Reading Outcome.} 
We measure differences in participants' social media news reading outcomes under two reading support conditions from three aspects.
\textbf{i) Comprehension.}
We measure participants' comprehension of the news by scoring their written artifacts using two items adapted from \cite{kovacs2014smart}
\textit{Understanding:} How much do the participants seem to understand the news overall? 
\textit{Consistency:} How consistent is the content in the participants' writing about the news? 
\textbf{ii) Critical thinking thoughts.} 
Based on \cite{ILLINOISTCriticalThinkingEssaytest}, we adapt four items to measure participants' critical thinking thoughts: 
\textit{Focus:} How well does the content demonstrate the participants' main ideas?
\textit{Diversity:} How well does the content demonstrate the breadth of the participants' analysis (\eg topics, perspectives)? 
\textit{Logic:} How well does the content demonstrate the clarity of the participants' analytical process? 
\textit{Solidity:} How well does the content demonstrate effective support for participants' viewpoints? 
For each participant's writing, we rate each of the above six items, which are rated on a five-point Likert Scale (1/5 - very bad/good). 
To ensure fairness in scoring, two researchers involved in article selection independently blind-score the participants' writing.
We average their scores as the final score for each item.
\textbf{iii) Self-perception.} Following \cite{collier2012conflict, liu2023coargue, wambsganss2021arguetutor}, we measure participants' self-perception of their writing from
\textit{Confidence:} How confident are the participants in their comprehension and critical thinking thoughts? 
\textit{Perceived quality:} How well do the participants feel about their comprehension and critical thinking thoughts?
After each stage, participants are required to complete a questionnaire self-assessing the two items above, using a seven-point Likert Scale (1/7 - very bad/good).

\textbf{RQ2: Reading Process.}
\textbf{i) Behaviors.}
To inspect participants' behavior during the reading process with \Toolname{}/Baseline, we log the
\cscwrr{
completion time of each reading session, writing session, the length of the user-written text, and the main button usage count.
}
\textbf{ii) Perceived engagement and task workload.}
We measure participants' perceived engagement in the reading process from seven aspects adapted from \cite{csikszentmihalyi1990flow, o2016theoretical}, \ie Concentration, Sense of Ecstasy, Clarity, Sense of Serenity, Doability, Timelessness Feeling, and Intrinsic Motivation. 
We also measure their perceived task workload using metrics from NASA Task Load Index \cite{colligan2015cognitive} regarding \ie Mental Demand, Physical Demand, Temporal Demand, Performance, Effort, and Frustration.

\textbf{RQ3: Perceptions towards the tool.} 
For each tool, we adopt the technology acceptance model from \cite{peng2022crebot, venkatesh2008technology, wambsganss2020adaptive} to measure its \textit{usefulness} (four items; Cronbach's \yuan{$\alpha=0.950$}), \textit{ease of use}  (four items; Cronbach's \yuan{$\alpha=0.948$}), and \textit{intent to use}  (two items; Cronbach's \yuan{$\alpha=0.954$}).
We average the scores of multiple items as the final score for each aspect. 
In addition, we ask for their opinions and suggestions on the tool regarding critical news reading assistant tools.

\kangyu{
\subsection{Complementary Study}
}
\subsubsection{Qualitative Study with Participants Who Do Not Often Read Comments}
To further learn about how those who do not often read comments perceive and use our system, we conduct a qualitative study with eight participants (P25-32) who self-reported that they read comments less than five times per ten news, \cscwfr{also from different disciplines such as artificial intelligence, education, and international politics}. 
Four of them are undergraduates, and the rest four are postgraduates. 
Five of them are males, and the other three are females. 
The mean age is $M = 23$, with $SD = 2.78$. 
Participants generally maintained daily social media news reading experience ($M = 1.44$ hours a day; $SD = 0.86$). 
They are required to first read an assigned piece of news on Twitter to arouse their memory on daily reading practice and then they read other two pieces of news with \Toolname{}. 
We prompt them to spend around five minutes on each reading task.
They are evenly and randomly allocated in one of the following conditions:  Policy (Twitter) - Health\&Religion (\Toolname{}), Gender (Twitter) - Health\&Religion (\Toolname{}), Health (Twitter) - Policy\&Gender (\Toolname{}), and Religion (Twitter) - Policy\&Gender (\Toolname{}). 
We don't require them to write down their thoughts about the news nor rate their perceptions with the reading process and tool. 
Instead, after all reading tasks, we conduct semi-structured interviews with each participant, with guiding questions about how \Toolname{} affected their reading behaviors and process (RQ2) and how they perceive \Toolname{} (RQ3). 

\kangyu{
\subsubsection{Case study on using \Toolname{} in daily social media news reading}
To explore how readers naturally perceive and use our system in daily social media news consumption, we conduct a case study with two participants (P33–P34). 
P33, a 22-year-old male undergraduate, reported reading self-media news for about 30 minutes daily; P34, a 24-year-old female postgraduate, averaged an hour. 
They browse four news following their usual habits, with no assigned tasks, and could freely decide whether and how to use \Toolname{}. 
Afterward, a semi-structured interview examines the system’s influence on their news understanding, critical thinking, and self-perception (RQ1), reading behaviors and processes (RQ2), and overall perceptions of \Toolname{} (RQ3).
}

\section{ANALYSES AND RESULTS}
To evaluate the differences in participants' comprehension and critical thinking with different reading conditions, we conducted a Wilcoxon signed-rank test \cite{woolson2007wilcoxon} to compare the performance of participants in each group.
The Wilcoxon signed-rank test is commonly used to compare two sets of scores that come from the same participants (\eg in HCI studies \cite{liu2022planhelper, peng2022crebot}).
\yuan{The test results indicate that users utilizing \Toolname{} demonstrated a higher level of comprehension and critical thinking compared to those using the baseline.}
In addition, we also perform the Wilcoxon signed-rank test \cite{woolson2007wilcoxon} to assess the difference in users' \textit{self-perception} in RQ1, as well as for the rated items in RQ2 and RQ3. 
For the participants’ comments and suggestions on the assistant tool, we perform a thematic analysis \cite{braun2012thematic}.
Two authors first coded all qualitative data independently and after discussion, then they formed a list of initial codes.
After several rounds of coding comparisons and discussions, they consolidate different codes into the pros and cons of each tool (Table \ref{pro_con}), which are incorporated into the result presentation below.

\subsection{RQ1. Reading Outcome}

\begin{figure}[htbp]
\centering
\includegraphics[width=\linewidth]{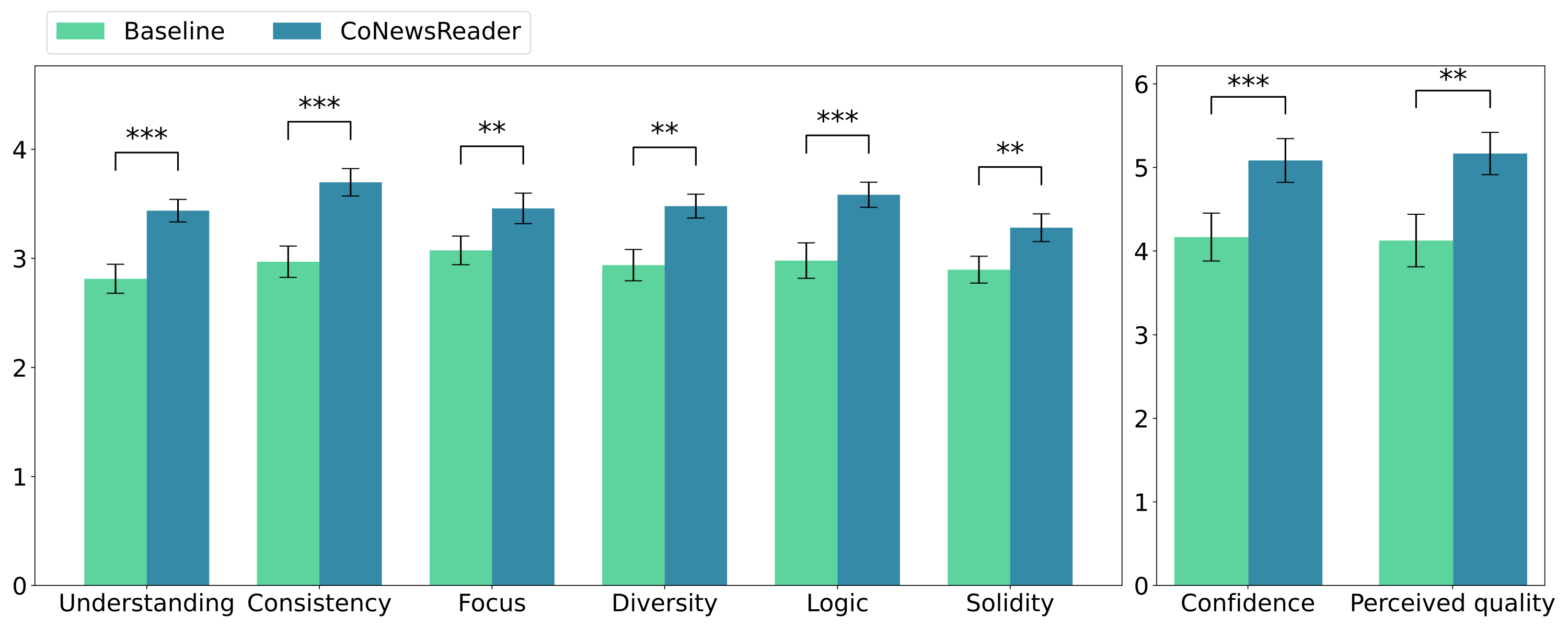}
\caption{(RQ1 results) The means, standard variation, and 95\% confidence intervals of the rated scores about participants’ writing after reading with \Toolname{} and Baseline, and participants' self-perception on their writing; *: p<.05; **:p<.01; ***:p<.001.}
\label{rq1figure}
\end{figure}

Figure \ref{rq1figure} shows the RQ1 results about \Toolname{}'s impact on users' critical reading outcomes in comparison to the Baseline.

\textbf{i) Comprehension.} %
Overall, participants with \Toolname{} performed better in terms of \textit{Understanding}  \yuan{(\Toolname{}: $M = 3.44$, $SD = 0.50$; Baseline: $M = 2.81$, $SD = 0.64$; $W = 14.5$, $p<.001$) and \textit{Consistency}  (\Toolname{}: $M = 3.70$, $SD = 0.60$; Baseline: $M = 2.97$, $SD = 0.69$; $W = 2$, $p<.001$)}, compared to the Baseline.
\yuan{15} participants explicitly stated that \Toolname{}'s main points and relevance identification helped them focus on reading the relevant information.
``\textit{With the key points of the news content summarised by \Toolname{}, I can get the relevant information quickly from a huge amount of information.}'' (P3, Male, age: 22). ``\textit{I like the fact that the tool can purify the information of the news, and it helps me to grasp the content of this news quickly.}''(P10, Female, age: 23). 

\textbf{ii) Critical thinking.}
The results indicate that with \Toolname{}, participants demonstrated better performance in \yuan{\textit{Focus} (\Toolname{}: $M = 3.46$, $SD = 0.67$; Baseline: $M = 2.97$, $SD = 0.63$; $W = 13.5$, $p<.01$), \textit{Diversity} (\Toolname{}: $M = 3.48$, $SD = 0.52$; Baseline: $M = 2.94$, $SD = 0.69$; $W = 39.5$, $p<.01$), \textit{Logic} (\Toolname{}: $M = 3.58$, $SD = 0.55$; Baseline: $M = 2.98$, $SD = 0.78$; $W = 7.5$, $p<.001$), and \textit{Solidity} (\Toolname{}: $M = 3.28$, $SD = 0.61$; Baseline: $M = 2.90$, $SD = 0.59$; $W = 15$, $p<.01$)} compared to using the Baseline.
\yuan{The main points and the categorised presentation of the comments enabled the participants to gain a deeper understanding of their own viewpoints and to consider points that they had not previously focused on.
``\textit{I like the main points and how they work with the comments. 
It helps me to check the comments against each other as I read, which makes it easier for me to understand the summary of the points.}'' (P23, Male, age: 23).
``\textit{
It's useful to have the comments organised into categories, as it shows me points that I hadn't noticed and helps me think about them.}'' (P18, Female, age: 25).
In addition, they could articulate their thoughts more clearly and extract more supporting information from comments to bolster their arguments.
``\textit{I attempted to approach the analysis from different keywords as entry points. They helped me achieve clearer thinking.}'' (P8, Female, age: 22).
``\textit{I appreciate the comment filter, as it allows me to focus on reading insightful comments and utilize them to support my ideas.}'' (P3, Male, age: 22).
``\textit{With keywords, I can easily find the direction for my thoughts.}'' (P6, Male, age: 19)}

\textbf{iii) Self-perception.}
The results indicate that with \Toolname{}, users exhibited higher levels of \textit{Confidence} \yuan{(\Toolname{}: $M = 5.08$, $SD = 1.26$; Baseline: $M = 4.17$, $SD = 1.37$; $W = 4.5$, $p<.001$) and \textit{Perceived quality} (\Toolname{}: $M = 5.17$, $SD = 1.21$; Baseline: $M = 4.13$, $SD = 1.51$; $W = 18$, $p<.01$)} in their writings, compared to using the baseline.
By concentrating on reading useful comments, users better utilize others' comments to support their own viewpoints and perspectives.
''\textit{I realize that my ideas are supported by other valid comments.}'' (P11, Female, age: 22). 
''\textit{I see more ideas from other people who agree with me, so I think I am thinking from the right perspective.}'' (P12, Female, age: 22).

\begin{table*}
\centering
\captionsetup{width=\textwidth}
\caption{Users' perceived engagement and workload (RQ2ii) in the training process as well as their perceptions (RQ3) towards the \Toolname{} or Baseline.}
\begin{tabular}{ccccccc} 
\hline
\multirow{2}{*}{Category}   & \multirow{2}{*}{Factor} & \Toolname{} & Baseline  & \multicolumn{3}{c}{Statistics}  \\ 
\cline{5-7}
                            &                         & Mean/S.D.                    & Mean/S.D. & W    & p        & Sig.          \\ 
\hline
\multirow{7}{*}{Engagement} & Concentration           & 5.96/0.84                    & 5.33/1.14 & 11  & 0.0109 & *             \\
                            & Sense of Ecstasy        & 6.25/1.05                    & 4.00/1.55 & 0    & 0.0001 & ***            \\
                            & Clarity                 & 5.75/1.05                    & 5.25/1.36 & 36    & 0.1661 &               \\
                            & Doability               & 4.58/1.80                    & 3.63/1.70 & 10.5  & 0.0076 & **             \\
                            & Sense of Serenity       & 5.54/1.22                    & 4.54/1.47 & 30.5    & 0.0154 & *             \\
                            & Timelessness Feeling    & 5.67/1.25                    & 4.92/1.63 & 19.5    & 0.0658 &               \\
                            & Intrinsic Motivation    & 5.75/0.97                    & 4.33/1.43 & 3.5    & 0.0008 & ***             \\ 
\hline
\multirow{6}{*}{Workload}   & Mental Demand           & 4.38/1.41                    & 5.42/1.22 & 16.5    & 0.0021  & **              \\
                            & Physical Demand         & 2.21/1.12                    & 2.75/1.36 & 4.5  & 0.0130 & *               \\
                            & temporal Demand         & 3.46/1.68                    & 4.38/1.75 & 38    & 0.0634 &               \\
                            & Performance             & 5.08/1.35                    & 3.88/1.30 & 33 & 0.0066  &   **            \\
                            & Effort                  & 4.17/1.57                    & 5.17/1.11 & 17  & 0.0077 &   **            \\
                            & Frustration             & 2.33/0.75                    & 3.92/1.63 & 3  & 0.0005 & ***             \\ 
\hline
\multirow{3}{*}{Acceptance} & Usefulness              & 6.05/0.94                    & 3.81/1.57 & 7  & 0.0001  & ***            \\
                            & Easy to Use             & 5.66/1.20                    & 3.67/1.52 & 16  & 0.0003 & ***             \\
                            & Intention to use        & 5.42/1.31                    & 3.42/1.74 & 27    & 0.0007 & ***             \\
\hline
\end{tabular}
\label{rq23table}
\end{table*}
\subsection{RQ2. Reading Process} \label{sec:rq2_findings}
\textbf{i) Behaviors.} 
In general, participants performed equally in both conditions in terms of time length in reading ($p>.05$) and writing ($p>.05$), and the number of words in their personal thoughts ($p>.05$).
\cscwrr{Figure in Appendix \ref{fig: buttonusagecount} counts clicked times of the buttons, showing participants made full tries of the features provided by each system. }
Even though \Toolname{} helps users filter out low-information comments, they spend more time contemplating.
``\textit{I feel that when using \Toolname{}, I spend more time delving into the details of comments rather than skimming}'' (P8, Female, age: 22).
 
In the qualitative study, seven participants who do not often read comments expressed a similar thought that \Toolname{} could 
increase their interest in continuing to explore and read comments.
``\textit{I like to use the comment filter. I can go directly to the category I want to explore, and when I read the comments of one category, I will naturally want to see others' thoughts in other categories.}'' (P30, male, age: 20) 
``\textit{The original comments section was a mess. I feel that by categorizing, I have a clearer understanding of the ideas from comments. The reading process is like building a network of opinions in my mind, which makes me feel immersed.}'' (P27, female, age:26).

\textbf{ii) Perceived engagement and workload.}
As for the seven items that measure perceived engagement during reading, there are significance in five of them, \yuan{\textit{Concentration} ($p<.05$), \textit{Sense of Ecstasy} ($p<.001$), \textit{Dobility} ($p<.01$), \textit{Sense of Serenity} ($p<.05$), and \textit{Intrinsic Motivation} ($p<.001$) respectively.}
With \Toolname{}, users can focus their attention on a few insightful understandings and analysis, reducing the distraction of low-information comments.
``\textit{I appreciate the comment filter, which allows me to stay focused on obtaining useful information. Compared to \Toolname{}, I feel that in the baseline, I sometimes spend more time searching for others' opinions than actually reading deeply, which makes me feel tired and frustrated.}'' (P4, Male, age: 22)
At the same time, participants with \Toolname{} show significant differences in \yuan{\textit{Mental Demand} ($p<.01$), \textit{Physical Demand} ($p<.05$), \textit{Performance} ($p<.01$), \textit{Effort} ($p<.05$), and \textit{Frustration} ($p<.001$) } in workload.
\yuan{The main points provided by \Toolname{} assist users in obtaining pertinent information with greater efficiency, thus reducing the time required to scan through the comments and the frustration that arises when there is a greater uniformity of content.
``\textit{I like to read the main points because there are so many social media comments these days that it's easy to get tired and annoyed when you read one after the other and don't see a different point of view}'' (P14, Female, age: 23).
``\textit{With the baseline, I sometimes struggle to find the information I need due to too many irrelevant comments, which even make me reluctant to continue reading}'' (P8, Female, age: 22)}. 

In the qualitative study with those who do not often read comments, participants first reported a variety of methods to read critically without \Toolname{}, such as identifying frequently appearing keywords in news content and comments, searching for relevant information on external websites, and discussing news with people around them. 
Nevertheless, after using \Toolname{}, all eight participants agreed that comments were helpful for critical reading, and \Toolname{} affected the way they looked at the comments. 
``\textit{When I read comments before, I mainly looked at people's emotions and rarely paid attention to content analysis.
This tool makes me pay more attention to the details in the content, such as the logic in the comments}'' (P31, male, age: 21). 
``\textit{When I read normally, I rarely realize the need to think critically unless it conflicts with my prior knowledge. However, the tool's classification of comments and the questions raised made me unconsciously enter the critical thinking process}'' (P25, male, age: 25).

\subsection{RQ3. Perceptions Towards the Tool}
\cscwrr{\textbf{Table \ref{rq23table}} shows participants' ratings on their \textbf{technology acceptance} of \Toolname{} and Baseline. }
We observed a significant differences in all items, with \yuan{\textit{usefulness} ($p<.001$), \textit{ease to use} ($p<.001$), and \textit{intention to use} ($p<.001$).}
Users perceive the integration of \Toolname{} as natural because of its intuitive design and straightforward operation.
``\textit{I like \Toolname{}'s clear interactions. If I had the option to enable or disable it freely, I think I would use it in my daily life.}'' (P2, Male, age: 22)
\cscwrr{As for \textbf{their comments} on \Toolname{} and Baseline (\textbf{Table \ref{pro_con}})}, participants felt that \Toolname{} could help them to 
quickly comprehend news (15), extract effective comments (12), think critically from different perspectives (14), and extract attitude from comments(9). 
``\textit{With\Toolname{}, I quickly grasped the main viewpoints of the news and comments. I could directly focus on the details of the comments I wanted to read. Additionally, the critical hints provided me with many directions for thinking and questions.}'' (P3, Male, age: 22). 
However, \Toolname{} could not provide varied, or user-defined filtering (9), or answer the critical thinking questions (9). 
For the baseline tool, participants favor its minimal interaction (12), simple interface (12), and familiar operation (9).
``\textit{When I read for relaxation, I still prefer the simple design of the baseline.}'' (P4, Male, age: 22)
Nevertheless, it can not filter the low-quality comments (15), and could be not conducive to critical thinking (10).

\begin{table}
\centering
\caption{(\yuan{RQ3 results) Summary of users’ comments about pros and cons of \Toolname{}/Baseline}}
\scalebox{0.85}{
\begin{tabular}{cll} 
\hline
\multicolumn{1}{l}{}         & \multicolumn{1}{c}{Pros (the number of participants who mention)}                                                                                                                           & \multicolumn{1}{c}{Cons}                                                                                                                                    \\ 
\hline
 \Toolname{} & \begin{tabular}[c]{@{}l@{}}Quickly comprehend news (15);\\Extract effective comments (12);\\Think critically from different perspectives (14);\\Extract attitude from comments (9)\end{tabular} & \begin{tabular}[c]{@{}l@{}}Could not provide varied, \\or user-defined filtering (9);\\Could not answer the critical \\thinking questions (9)\end{tabular}  \\ 
\hline
Baseline                     & \begin{tabular}[c]{@{}l@{}}Minimal interaction (12); \\Simple interface (12):\\Familiar operation (9)\end{tabular}                                                                            & \begin{tabular}[c]{@{}l@{}}Can not filter the low-quality \\comments (15); Could be not \\conducive to critical thinking (10)\end{tabular}                    \\
\hline
\end{tabular}}
\label{pro_con}
\end{table}


In the qualitative study with those who do not often read comments, 
seven participants said they enjoyed using the tool and felt immersed in reading, but they said they might only use it when they wanted to obtain information and do in-depth reading. 
Three participants said that although the tool could filter out useful information, it might lack entertainment and fun. 
``\textit{Although entertaining content is not useful for critical thinking, it can provide some emotional value in daily life. After using the tool, the reading process becomes monotonous and serious, so I probably would not use it in daily life unless I am doing critical reading}'' (P29, male, age: 20).
Regarding the cognitive burden after using it, two participants said that it would be a little heavier than usual, but it was acceptable for the critical thinking process.
``\textit{I feel that I was more involved in the critical reading process, so the burden will be a little heavier than when I did not use this tool}'' (P30, male, age: 20).
\cscwrr{
At the same time, three participants (P26, P27, P32) noted that they could occasionally sense potential hallucinations in the model-generated content, but the explicit links back to original comments and news enabled them to make their own judgments, leading to general acceptance of the system.
``Sometimes I had a feeling the summary might slightly mismatch, but because I could always click back to the original comments, I could judge for myself what to trust. That made me feel comfortable relying on the system overall.'' (P27, female, age:26)
}

\kangyu{
In naturalistic, task-free use, both participants integrated \Toolname{} into their daily news reading, adjusting engagement depth based on personal interest. 
After they read through a post's content, they typically begin with the system’s main points rather than directly exploring comments, enabling a rapid grasp of the incident, especially for unfamiliar topics. 
They often navigated from a point to its source comments for verification and context, shifting from fragmented skimming to more structured, iterative exploration. 
After that, the comment attitude filter was usually used to compare stances, quickly assess viewpoints, and surface minority perspectives otherwise buried in traditional ranked lists, while the comment content filter was selectively applied to personally interesting topics to locate comments of higher value for reading or thought-provoking. 
Critical hints were likewise used on high-interest or contentious topics, prompting consideration of perspectives not previously attended to, even if only contemplated briefly. 
For highly interesting topics, participants engaged more comprehensively—reviewing all main points, exploring linked comments, and applying multiple filters—whereas for less engaging topics, interaction was often limited to the post, main points, and comment attitude filter. 
Overall, two participants regarded the system as a lightweight, on-demand scaffold that encouraged deeper comment exploration and integrated diverse perspectives, without disrupting habitual reading practices.
}

\cscwrr{
\subsection{Real-world User Scenarios}
To better demonstrate user scenarios in the real world, in this case, we describe in detail how P33 uses CoNewsReader for reading important news during his everyday social media use.  
He comes across a trending post about a new policy that “allows transgender people to use the gym locker room that matches their self‑identified gender” and, out of curiosity, quickly browses the post.  
Finding the wording overly emotional and the background information insufficient, he decides to invoke CoNewsReader.  
P33 first checks the system’s main points to obtain an overview of the incident, then clicks on several points that summarize key controversies and navigates to the corresponding comments.  
By iteratively examining these summaries and comments, he reports that this “deepened his understanding of the stakeholders involved and the concrete details of the incident.”  
He then uses the comment attitude filter to compare supportive and opposing stances, and applies the comment content filter to surface more analytical comments.  
These diverse analytical perspectives prompt him to reconsider his initially one‑sided impression and to reflect on the issue from different angles, such as safety, children, and privacy.  
He briefly reviews the critical hints and notices one critical thinking question: “How can it ensure the safety and comfort of all patrons?”  
This leads him to consider alternative solutions, such as adding an additional locker room for transgender users rather than completely changing the existing layout.  
Overall, after reading this post with CoNewsReader, P33 feels that he has reduced the emotional reactions triggered by his first impression, gained a clearer understanding of the incident, and become better able to analyze the pros and cons from multiple perspectives.
}
\section{Discussion}
In this work, we propose \Toolname{} to assist users in comprehending and raising critical thinking thoughts while reading news on social media. 
In our user study, we found that compared to reading the news with baseline tool, participants with \Toolname{} demonstrated significantly better performance \yuan{in the focus, diversity, logic, and solidity of their thoughts.}
As commented by our participants, this can be accounted for by the comments with other readers' \textit{Skepticism} and \textit{Provocation} and critical thinking keywords and questions of \Toolname{}.
In line with previous work \cite{yuan2023critrainer, chen2023marvista, petridis2023anglekindling}, our findings also support the results that the specific guidance adapted from the text could facilitate the critical thinking process.

\cscwrr{As \Toolname{} is fundamentally comment‑driven, its effectiveness depends on the depth and volume of user contributions. 
It works more suitably for consequential, knowledge‑intensive or contested news that attracts richer, more reflective comments, and may be less effective for lightweight or entertainment‑oriented news with shallow or sparse comments.
}
\textbf{Our ideas of \Toolname{} can also be applied to assist users' critical thinking in other scenarios}, \eg \cscwrr{knowledge community like reddit,} short video platforms like TikTok, shopping platforms like Amazon, or learning platforms like MOOC with user-generated content (\eg comments) and machine-generated content (\eg summary and critical thinking questions). 
For example, \Toolname{} could summarize main points from short videos, highlight critical comments, and generate thought-provoking questions, helping users think more efficiently, diversely, and critically while mitigating bias and cognitive tendencies \cite{yuan2023critrainer, chen2023marvista}.
\cscwrr{
In such settings, our workflow could still organize and contrast user contributions from multiple angles to support critical thinking, but the “news‑event summarization” function of \Toolname{} would be less applicable, as there is no single underlying article or event to summarize.
}
\textbf{At a broader level, \Toolname{} supports and can be generalized to a new news-reading paradigm}, \ie instead of reading one news, people are able to simultaneously browse a large set of information beyond a single article. 
\Toolname{} distills information from comments and aggregates it into the summary of a news, categorizes community opinions, and offers critical hints users should pay attention to. 
\kangyu{Stepping forward, with similar features, \Toolname{} could easily support lateral reading \cite{zhang2023readprobe}, that users can read multiple related news with their comments at the same time, to better evaluate online information.} 
Expanding this approach increases both its influence and responsibility, highlighting the need to collaborate with stakeholders—such as readers, publishers, and creators to develop guidelines for fair use.
For instance, while readers may appreciate diverse content, creators may worry about diluted viewpoints they intend to convey.
Despite different interests, as enacted in \Toolname{}, a key guideline should be that summary and critical questions are grounded in human-contributed content, not purely AI-generated text. 
\textbf{In addition, tools like \Toolname{} could support participatory journalism} \cite{bowman2003we}, in which citizens take an active role in collecting, reporting, analyzing, and disseminating news and information. 
By organizing and summarizing perspectives and factual information from user comments and news content,\Toolname{} may facilitate broader audience engagement in news discussion and interpretation, as well as help enhance the quality of user-generated insights. 
This could lower barriers to public participation and help surface a wider range of viewpoints within news ecosystems. 
\cscwrr{
With a new self-expression as the fourth stage, such tools may have the potential to foster a more inclusive and interactive journalistic environment \cite{liu2023coargue}, enabling audiences to play a more active role in the construction and dissemination of online news.
}
\kangyu{
\textbf{\Toolname{} is not intended to replace platforms such as Medium or other in-depth news sources}, but to support a complementary scenario that reading news on social media. 
It addresses key limitations of social media news—such as missing context, hard-to-access critical-target comments—by providing a lightweight tool that fosters critical thinking during in-platform engagement. 
When deeper information is needed, users can transition to external sources, while in cases where authoritative news is scarce for fragmented we-media news \cite{juneja2022human, wang2021journalistic}, the system offers supplemental cues and perspectives to support a more comprehensive understanding.
}

\subsection{Design Considerations}

From our design and evaluation study of \Toolname{}, we derive design considerations for future critical news reading support tools from the aspects of features, models, and interaction design. 

\subsubsection{Personalize critical news reading with comments}
One unique feature in \Toolname{} is that it generates a bullet-point summary of news combined with information-related comments, which is different from the previous news summarization tools that only rely on the content of the news \cite{chen2023marvista, petridis2023anglekindling, bhuiyan2023newscomp}. 
Based on findings about RQ1i, this summary helps participants comprehend the news better and enables them to focus on reading the relevant information of their interests. 
Feedback from participants (Table \ref{pro_con}) also indicates that our comment filters help users extract helpful comments that foster critical thinking.
Therefore, we suggest future news-reading support tools to make use of the comments to facilitate personalized news reading. 
Our \Toolname{} offers three such example features to get started, \ie comment-enhanced news summary, comment filters, and comment-driven critical thinking questions. 
Our participants further suggested that these tools could provide varied or user-defined filtering of comments (Table \ref{pro_con}). 
Future tools could explore such suggested features, \eg allowing users to input keywords to further filter comments within a comment group and enabling users to mark any comment of their interest and get a news summary accordingly \cite{wu2024comviewer}.

\subsubsection{Model user preferences of helpful comments for news reading}
The key features of \Toolname{} largely rely on the classifications of news comments that are helpful for critical news reading. 
We demonstrated in \autoref{sec:comments_types} that we collected and categorized a small sample of useful and non-useful comments with twelve participants, in \autoref{binaryclassifiers} that we prompted LLMs to classify these comments, and in the user study that the features built on these classified comments fostered critical news reading. 
Our approach for modeling comments is lightweight in terms of dataset and developed model, and it could serve as a good starting point for future news-reading support tools that would like to satisfy user needs for other types of helpful comments. 
For example, the tools could allow users to mark a few samples (\eg 20 for each class) of comments they like during their daily reading activities and adopt the approach similar to ours to classify that class of comments of news.

\subsubsection{Align interaction design of the tool with everyday news reading habits}
In \Toolname{}, we structure a 3R critical reading framework (\autoref{sec:3r}) that nudges users to read the news content and comments and make reflections. 
Correspondingly, \Toolname{} has three new features added to the traditional social media interfaces, \ie news summarization, comment filters, and critical questions (\autoref{Comparison}). 
While these features ease users' workload in the critical news reading tasks (\autoref{rq23table}), they complicate the interface and interaction design of the tool (\autoref{pro_con}). 
Therefore, we suggest future tools to explore interaction design that better aligns with users' everyday news reading habits. 
For example, some users are used to highlighting text when reading, which could help reading comprehension \cite{joshi2024constrained}, and a tool like \Toolname{} could pop up the critical thinking questions next to the selected or highlighted text.

\subsection{Concerns}
\subsubsection{Misinformation, Biases, and Hallucinations}
\Toolname{} can incorporate comments into a summary of news in the \cscwfr{Read content} stage and leverage comments to generate critical thinking questions in the reflection stage. 
While our user study demonstrates that these summaries and questions are helpful for supporting critical news reading, we should be aware of the risks associated with our approach. 
\cscwrr{First}, the user comments may convey inaccurate or fake information, which would misguide the news summary and affect the readers' understanding \cite{hsueh2015leave}.
\cscwrr{Second}, the user comments may convey biased opinions, which would result in critical thinking questions that bias readers' thoughts. 
\cscwrr{Third, similar to other LLM‑based systems, our model may occasionally generate hallucinations, such as slight mismatches between summaries and the linked comments.}

To mitigate the potential risks of \Toolname{}, future work can explore approaches before, during, and after incorporating comments in the generation tasks. 
For example, in addition to our binary classifiers for identifying each type of useful comments for critical news reading (\autoref{classification}), we could further invoke models for fact-checking and bias detection \cite{mu2023self, portelli2020distilling} to filter comments as input to the generation tasks. 
We could also control the sampling ratio from comments expressing different opinions and attitudes when generating summaries or questions to mitigate bias \cite{dai2024bias}. 
\cscwrr{In addition, future work can explore reducing the possibility of model generation bias and hallucinations in our critical news reading support task by further training or fine-tuning the generative models (\eg by introducing expert-annotated news datasets that encode nuanced degrees of factuality and reliability) \cite{rashkin2017truth, cheng2021mitigating, mozafari2020hate}. }
Moreover, future researchers could prompt the models to generate multiple different opinions for a keyword and display them simultaneously, such as summarizing different branching angles for news in \cite{petridis2023anglekindling}. 
After all, in case the generated summary and questions still contain misinformation or biases, tools like \Toolname{} should remind readers of potential risks attached to the generated content, \eg ``the content is generated by AI and may contain inappropriate information''. 
\kangyu{
Further, readers could also be encouraged to maintain a critical mindset at all times—whether toward machine-generated content, the news itself, or user-generated comments \cite{lim2025iffy}.
}

\subsubsection{Echo Chamber Effect}
The echo chamber effect refers to a situation in which information, ideas, or beliefs are amplified and reinforced by communication and repetition inside a closed system, which may lead to a distorted perception of reality. 
Our \Toolname{} could reduce such an effect by helping readers pay attention to the comments that contain event information, express viewpoints, or doubt the news, rather than the popular comments that often echo mainstream thoughts. 
This benefit is supported by the results in \autoref{rq1figure} that the diversity of thinking angles in participants' writing was significantly improved when using our system. 
However, there is also a risk of reinforcing the echo chamber effect if the comments under a news were not diverse enough, especially when many comments agree with each other and form a strong viewpoint together. 
In addition, with our filters, readers could only choose the type of comments they want to read, which may also potentially enhance the echo chamber effect. 
In these cases, language agents with personas sourced from community data \cite{zhang2024see}could help mitigate the echo chamber effect by contributing more viewpoints and encouraging users to discuss these viewpoints.

\cscwrr{
\subsubsection{Risks and Challenges in Scaling}
Deploying \Toolname{} at scale may expose it to a broader range of adversarial and structural risks. 
Because the system is comment‑driven, coordinated groups of users or bots could flood discussions with aligned viewpoints or misleading narratives that our summarization and highlighting may inadvertently amplify (a form of collusion attack) \cite{ratkiewicz2011detecting}.
Adversaries might also game our ranking and filtering by crafting comments that superficially appear “critical,” “informative,” or “balanced,” while still promoting a particular agenda, or by using targeted brigading and harassment to push hostile or inflammatory remarks to the foreground \cite{luca2016fake}. 
In addition, attackers could strategically manipulate agent‑bots’ personas or prompts (e.g., by seeding biased “community norms”) to steer how viewpoints are generated and framed \cite{mayzlin2014promotional}. 
More broadly, domain shifts across platforms, languages, and cultural contexts may cause models trained on one community’s data to misclassify or under‑represent perspectives in another \cite{del2020linguistic}. 
While systematically addressing these threats is beyond the scope of our current prototype, they highlight the need for future work on adversarially robust comment selection \cite{bai2021recent}, detection of coordinated or inauthentic behavior \cite{cinelli2022coordinated}, cross‑community generalization \cite{schaffner2024community, chandrasekharan2019crossmod}, and human‑in‑the‑loop moderation and auditing \cite{lai2022human}.
}

\subsection{Limitations and Future Work} \label{sec:limitations}
Our work presents several limitations.

\textbf{Study Design and Participant Diversity.} 
Our participants are university students from mainland China reading in their second language (English), limiting generalizability across different educational backgrounds, ages, professions, and cultural contexts.
\kangyu{
Social media news consumers with low media literacy or without prior critical thinking training are prone to being misled by fake news. 
They often rely on a first impression of the content, lack the initiative to retrieve verifiable facts, and have difficulty evaluating articles with unclear positions or those requiring domain-specific background knowledge \cite{simko2021study}. 
Future designs may consider incorporating more explicit and proactive cues—for example, as demonstrated by \citet{lim2025iffy}, who automatically flagged fallacies and provided contextual information. 
}
Additionally, our experimental design has several constraints: we simplified social media features (removing hierarchical comment threads and user profiles) to focus more on designed features, had news topics co-occur in ways that could influence results, and could not fully eliminate observer effects despite counterbalancing.
Future research should examine CoNewsReader's effectiveness with diverse user populations and implement more rigorous blinding procedures to mitigate potential biases.

Our laboratory-based evaluation with specific writing tasks may not reflect everyday news-reading behaviors. 
For real-world impact, \Toolname{} would need adaptation to different contexts (\eg mobile reading during commutes) and time constraints (\eg brief reading sessions under three minutes). 
Future implementations could incorporate adaptive support based on available reading time, similar to \citet{chen2023marvista}'s approach, to better accommodate varied reading situations while still encouraging critical engagement.

\textbf{Comment Classification Refinement.} 
Our taxonomy of CNR-functional comments, while grounded in user perspectives, is limited by our sample and selected news. 
Alternative approaches to identifying valuable comments could emerge from different domains (\eg health informatics) or considering additional factors (\eg bot-generated content). 
Expanding this taxonomy with input from diverse users with different reading habits and concerning different news topics could yield a more comprehensive classification of helpful comments.

\textbf{Multimodal Content Processing.}
Currently, CoNewsReader processes only text, despite the increasingly multimodal nature of news and comments on social media. Incorporating multimodal language models like Kosmos-2 \cite{peng2023kosmos} would allow the system to process images and videos, providing more comprehensive support for critical reading of contemporary media content.

\textbf{Alternative Framework Approaches.} 
Our implementation of the 3R framework presents critical thinking prompts after content consumption, but alternative timing and presentation approaches merit exploration.
Future systems could experiment with in-situ questioning during reading \cite{peng2022crebot}, user-generated questioning based on templates \cite{yuan2023critrainer}, or inverting the reading order (comments before articles) to investigate different effects on critical engagement.
These approaches might offer additional benefits for supporting critical news consumption in varied contexts.

\cscwrr{
\textbf{Long Comments Processing.}
Our current implementation is optimized for typical social media comments of short to moderate length, and we have not systematically evaluated \Toolname{} on very long, multi-paragraph site comments or long-form user posts. This creates a potential limitation: the prompt templates and interaction flow may not scale well when individual comments approach article-level length, potentially affecting both performance (due to context limits) and the granularity with which user perspectives are represented. Future work could consider incorporating techniques such as long-text segmentation \cite{cho2022toward} and hierarchical summarization \cite{flokas2025towards} to better handle substantially longer commentary. These extensions could enable the system to preserve more nuanced arguments from long comments while still providing concise, actionable support for critical news reading.
}

\section{Conclusion}
In this paper, we design and build a comment-powered assistant tool \Toolname{} to support users in comprehending and raising critical thoughts during the critical reading process on social media. 
\Toolname{} can classify news comments based on content and sentiment. 
It summarizes and provides relevance identification based on information-related comments to support comprehending the news, and utilizes inspiring comments to generate keywords and critical thinking questions to support raising critical thinking thoughts.
The comment filter provided by \Toolname{} performs well to help users find useful comments quickly, reducing the time and burden of finding information. 
We compared \Toolname{} with the current reading condition Twitter-like Baseline through a within-subjects study with \yuan{24} participants who often read comments. 
The Results show that \Toolname{} can not only better facilitate participants' comprehension of news, but also better help users raise more focused, diverse, logical, and solid critical thinking thoughts. 
\revise{The qualitative study with eight participants who do not often read comments and \kangyu{case study with two participants in a daily social media news reading scenario} show that \Toolname{} can increase participants' interest in exploring comments to potentially help with critical thinking.}
Our work offers insights and design considerations for building intelligent tools to assist in comprehending and critical thinking based on user- and machine-generated content.
\begin{acks}
This work is supported by the General Projects Fund of the Natural Science Foundation of Guangdong Province in China grant (2024A1515012226). We are grateful to the anonymous reviewers for their valuable feedback and to our participants for their crucial contributions to advancing this research.
\end{acks}


\bibliographystyle{ACM-Reference-Format}
\bibliography{main}
\appendix
\newpage
\section{Evaluation rubrics}
\label{rubrics}

\begin{table}[htbp]
\footnotesize
\caption{Evaluation rubrics for news summarizer}
\begin{tabular}{|p{1cm}|p{0.17\textwidth}|p{0.17\textwidth}|p{0.17\textwidth}|p{0.17\textwidth}|p{0.17\textwidth}|}
\hline
 &
  \multicolumn{5}{c|}{\textbf{News summarizer}} \\ \hline
 &
  \multicolumn{1}{c|}{\textit{Readability}} &
  \multicolumn{1}{c|}{\textit{Completeness}} &
  \multicolumn{1}{c|}{\textit{Relevance}} &
  \multicolumn{1}{c|}{\textit{Precision}} &
  \multicolumn{1}{c|}{\textit{Trustworthiness}} \\ \hline
\textbf{Score 1} &
  \begin{tabular}[c]{@{}p{0.17\textwidth}@{}}The summary is very difficult to understand, with numerous grammatical errors and incoherent sentence structures. Key ideas are obscured by poor language use. \end{tabular} &
  \begin{tabular}[c]{@{}p{0.17\textwidth}@{}}The summary covers very little of the original content,  missing most of the main points and details.\end{tabular} &
  \begin{tabular}[c]{@{}p{0.17\textwidth}@{}}The citation serial number is completely irrelevant to the main points of the original text,  providing no useful connection or context.\end{tabular} &
  \begin{tabular}[c]{@{}p{0.17\textwidth}@{}}The key points are extremely vague, lacking specific information and clarity,  making it difficult to understand the intended message.\end{tabular} &
  \begin{tabular}[c]{@{}p{0.17\textwidth}@{}}The key points are highly untrustworthy, containing significant inaccuracies or misleading information that contradicts the original news and comments.\end{tabular} \\ \hline
  
\textbf{Score 2} &
  \begin{tabular}[c]{@{}p{0.17\textwidth}@{}}The summary is somewhat challenging to understand, with several grammatical errors and awkward phrasing that distracts from the content. \end{tabular} &
  \begin{tabular}[c]{@{}p{0.17\textwidth}@{}}The summary covers some parts of the original content and omits several key points and important details.\end{tabular} &
  \begin{tabular}[c]{@{}p{0.17\textwidth}@{}}The citation serial number is somewhat related  to the original text but misses several key points,  leading to confusion about its relevance.\end{tabular} &
  \begin{tabular}[c]{@{}p{0.17\textwidth}@{}}The key points are somewhat imprecise,  containing general statements that do not clearly convey the main ideas or details.\end{tabular} &
  \begin{tabular}[c]{@{}p{0.17\textwidth}@{}}The key points are somewhat untrustworthy,  including several inaccuracies or misrepresentations of the original content.\end{tabular} \\ \hline
  
\textbf{Score 3} &
  \begin{tabular}[c]{@{}p{0.17\textwidth}@{}}The summary is moderately understandable, with some grammatical errors and unclear sentences, but the main ideas can still be discerned. \end{tabular}&
  \begin{tabular}[c]{@{}p{0.17\textwidth}@{}}The summary covers about half of the original content,  including some key points but missing other significant details.\end{tabular} &
  \begin{tabular}[c]{@{}p{0.17\textwidth}@{}}The citation serial number is moderately relevant,  linking to some main points of the original text but not fully capturing the overall context.\end{tabular} &
  \begin{tabular}[c]{@{}p{0.17\textwidth}@{}}The key points are moderately precise,  including some specific information but still  leaving room for ambiguity in certain areas.\end{tabular} &
  \begin{tabular}[c]{@{}p{0.17\textwidth}@{}}The key points are moderately trustworthy,  with some accurate information but also contain a few inaccuracies and vague interpretations of the original news and comments.\end{tabular} \\ \hline
  
\textbf{Score 4} &
  \begin{tabular}[c]{@{}p{0.17\textwidth}@{}}The summary is mostly clear and easy to read, with minor grammatical errors and generally coherent sentence structures. \end{tabular}&
  \begin{tabular}[c]{@{}p{0.17\textwidth}@{}}The summary covers most of the original content,  capturing the main points but leaving out minor details.\end{tabular} &
  \begin{tabular}[c]{@{}p{0.17\textwidth}@{}}The citation serial number is mostly relevant,  corresponding well to the main points of the original text and providing a clear connection.\end{tabular} &
  \begin{tabular}[c]{@{}p{0.17\textwidth}@{}}The key points are mostly precise,  providing clear and specific information that effectively conveys the main ideas with minor vagueness.\end{tabular} &
  \begin{tabular}[c]{@{}p{0.17\textwidth}@{}}The key points are mostly trustworthy,  accurately reflecting the original news and comments with only minor discrepancies.\end{tabular} \\ \hline
  
\textbf{Score 5} &
  \begin{tabular}[c]{@{}p{0.17\textwidth}@{}}The summary is very clear, completely free from grammatical errors, and highly coherent, making it very easy to read and understand. \end{tabular} &
  \begin{tabular}[c]{@{}p{0.17\textwidth}@{}}The summary fully covers the original content,  capturing all key points and relevant details comprehensively.\end{tabular} &
  \begin{tabular}[c]{@{}p{0.17\textwidth}@{}}The citation serial number is highly relevant,  directly corresponding to the main points of the original text and enhancing the overall understanding.\end{tabular} &
  \begin{tabular}[c]{@{}p{0.17\textwidth}@{}}The key points are highly precise, containing clear,  specific information that accurately conveys the main ideas without any vagueness.\end{tabular} &
  \begin{tabular}[c]{@{}p{0.17\textwidth}@{}}The key points are highly trustworthy,  completely aligning with the original news and comments and accurately representing the information without any misleading elements.\end{tabular} \\ \hline
\end{tabular}
\label{summarizer_rubrics}
\end{table}

\begin{table}[htbp]
\footnotesize
\caption{Evaluation rubrics for keyword extractor \& question generator}
\begin{tabular}{|p{1cm}|p{0.3\textwidth}|p{0.3\textwidth}|p{0.3\textwidth}|}
\hline
 &
 \multicolumn{3}{c|}{\textbf{\begin{tabular}[c]{@{}c@{}}keyword extractor \& question generator\end{tabular}}} \\ \hline
 &
 \multicolumn{1}{c|}{\textit{Relevance}} &
 \multicolumn{1}{c|}{\textit{Accessibility}} &
 \multicolumn{1}{c|}{\textit{Usefulness}} \\ \hline
\textbf{Score 1} &
 \begin{tabular}[c]{@{}p{0.3\textwidth}@{}}The keywords are completely unrelated to the original content, providing no useful connection or context.\end{tabular} &
 \begin{tabular}[c]{@{}p{0.3\textwidth}@{}}The questions are very difficult to understand, filled with complex language and unclear phrasing, making them confusing and inaccessible.\end{tabular} &
  \begin{tabular}[c]{@{}p{0.3\textwidth}@{}}The questions are not useful at all, failing to stimulate any further critical thinking or personal reflection.\end{tabular} \\ \hline
 
\textbf{Score 2} &
 \begin{tabular}[c]{@{}p{0.3\textwidth}@{}}The keywords are somewhat related to the original content but miss several key themes, leading to confusion about their relevance.\end{tabular} &
 \begin{tabular}[c]{@{}p{0.3\textwidth}@{}}The questions are somewhat challenging to understand, with awkward phrasing and several unclear terms that detract from their clarity.\end{tabular} &
  \begin{tabular}[c]{@{}p{0.3\textwidth}@{}}The questions are somewhat useful, but they only generate limited critical thinking and do not encourage deeper exploration of the topic.\end{tabular} \\ \hline
 
\textbf{Score 3} &
 \begin{tabular}[c]{@{}p{0.3\textwidth}@{}}The keywords are moderately relevant, linking to some aspects of the original content but not fully capturing the overall themes.\end{tabular} &
 \begin{tabular}[c]{@{}p{0.3\textwidth}@{}}The questions are moderately understandable, with some clear elements but also contain vague language or complex structures that may confuse some readers.\end{tabular} &
  \begin{tabular}[c]{@{}p{0.3\textwidth}@{}}The questions are moderately useful, prompting some critical thinking but lacking depth to fully engage personal reflection.\end{tabular} \\ \hline
 
\textbf{Score 4} &
 \begin{tabular}[c]{@{}p{0.3\textwidth}@{}}The keywords are mostly relevant, corresponding well to the main themes of the original content and providing a clear connection.\end{tabular} &
 \begin{tabular}[c]{@{}p{0.3\textwidth}@{}}The questions are mostly clear and easy to understand, using straightforward language and coherent structure, with only minor issues.\end{tabular} &
  \begin{tabular}[c]{@{}p{0.3\textwidth}@{}}The questions are mostly useful, effectively encouraging further critical thinking and prompting deeper personal insights about the topic.\end{tabular} \\ \hline
 
\textbf{Score 5} &
 \begin{tabular}[c]{@{}p{0.3\textwidth}@{}}The keywords are highly relevant, directly corresponding to the main themes of the original content and enhancing the overall understanding.\end{tabular} &
 \begin{tabular}[c]{@{}p{0.3\textwidth}@{}}The questions are very clear, completely free from ambiguity, and highly coherent, making them easy to understand and engage with.\end{tabular} &
  \begin{tabular}[c]{@{}p{0.3\textwidth}@{}}The questions are highly useful, significantly stimulating further critical thinking and inspiring deep personal reflection and exploration of the topic.\end{tabular} \\ \hline
\end{tabular}
\label{keyword_rubrics}
\end{table}
\FloatBarrier
\cscwrr{
\section{Button usage count}
\label{fig: buttonusagecount}
\begin{figure}[htbp]
    \centering
    \includegraphics[width=0.9\linewidth]{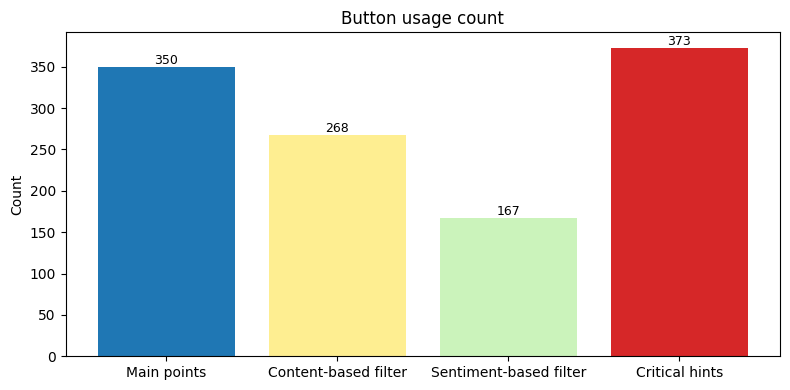}
    \caption{Usage frequency of main interface buttons in the between-subjects study with 24 participants. The button in Figure \ref{interface} b1 has been grouped into a content-based filter. The button in Figure \ref{interface} b2 has been grouped into a sentiment-based filter.}
\end{figure}
}
\end{document}